\documentclass[preprint,3p,onecolumn,number,11pt,margin=1in]{elsarticle}

\usepackage{graphicx}
\usepackage{amssymb}

\usepackage{lineno}

\usepackage{lineno,hyperref}
\modulolinenumbers[5]
\usepackage[latin1]{inputenc}
\usepackage{amsmath,empheq}
\usepackage{amsfonts}
\usepackage{amssymb}
\usepackage{graphicx}
\biboptions{super,comma}

\usepackage[font=small,format=plain,labelfont=bf,up,textfont=normal,up,justification=justified,singlelinecheck=false]{caption}
\usepackage{subfig}
\usepackage{float}
\usepackage{longtable}
\usepackage{color}
\usepackage{booktabs}

\usepackage{makecell}
\setcellgapes{5pt}

\usepackage[version=4]{mhchem}
\usepackage{verbatim}
\usepackage{epsfig} 
\usepackage{cleveref}
\graphicspath{{./Figures/}}
\usepackage{soul}

\usepackage{xr}
\usepackage{comment}
\usepackage{multirow}
\usepackage{lineno}
\usepackage{algpseudocode}
\usepackage{setspace}

\usepackage[normalem]{ulem}
\makeatletter
\def\ps@pprintTitle{%
\let\@oddhead\@empty
\let\@evenhead\@empty
\def\@oddfoot{\hfil\thepage\hfil}
\let\@evenfoot\@oddfoot
}
\makeatother

\begin{document}

\begin{frontmatter}

\title{Optimized tandem catalyst patterning for CO$_2$ reduction flow reactors}



\author[address1]{Jack Guo\corref{mycorrespondingauthor}}
\cortext[mycorrespondingauthor]{Corresponding author}
\ead{guo9@llnl.gov}

\author[address1]{Thomas Roy}

\author[address2]{Nitish Govindarajan}

\author[address2]{Joel B. Varley}

\author[address3,address5]{Jonathan Raisin}

\author[address3,address4,address6]{Jinyoung Lee}

\author[address6,address7]{Ji-Wook Jang}

\author[address3,address4]{Dong Un Lee}

\author[address3,address4]{Thomas F. Jaramillo}

\author[address1]{Tiras Y. Lin\corref{mycorrespondingauthor}}
\ead{lin46@llnl.gov}

\address[address1]{Computational Engineering Division, Lawrence Livermore National Laboratory, Livermore, CA 94550, USA}

\address[address2]{Materials Science Division, Lawrence Livermore National Laboratory, Livermore, CA 94550, USA}

\address[address3]{SUNCAT Center for Interface Science and Catalysis, Department of Chemical Engineering, Stanford University, Stanford, CA 94305, USA}

\address[address4]{SUNCAT Center for Interface Science and Catalysis, SLAC National Accelerator Laboratory, Menlo Park, CA 94025, USA}

\address[address5]{TotalEnergies Research \& Technology USA LLC, Houston, TX 77002, USA}

\address[address6]{School of Energy and Chemical Engineering, Ulsan National Institute of Science and Technology (UNIST), Ulsan, Ulsan, 44919, Republic of Korea}

\address[address7]{Graduate School of Carbon Neutrality, Ulsan National Institute of Science and Technology (UNIST), Ulsan, Ulsan, 44919, Republic of Korea}

\begin{abstract}
Tandem catalysis involves two or more catalysts arranged in proximity within a single reaction vessel, with the aim of synergistically aligning the catalysts' reaction pathways to maximize overall system performance.
This study presents a proof of concept showing the integration of continuum transport modeling with design optimization in a simplified two-dimensional flow reactor setup for electrochemical \ce{CO_2} reduction. \ce{Ag} catalysts provide the \ce{CO_2 -> CO} reaction capability, and \ce{Cu} catalysts provide the \ce{CO -> } high-value products reaction capability. Given a set of input parameters, the optimization algorithm uses adjoint methods to modify the \ce{Ag}/\ce{Cu} surface patterning in order to maximize the current density toward high-value products, such as ethylene. The optimized designs yield significant performance enhancement especially at more negative applied voltages (i.e., stronger surface reactions) and for larger numbers of patterning sections. For an applied voltage of $-1.7$ V vs. SHE, the $12$-section optimized design increases the current density towards ethylene by up to $65$\% compared to the unoptimized $2$-section design. 
For the optimized cases, observed differences in the production and consumption of \ce{CO} (the key intermediate species) and minimized zones of low \ce{CO} reactant surface concentration on \ce{Cu} sections explain the improved reactor performance.

\end{abstract}


\end{frontmatter}


\section{Introduction}

The electrochemical reduction of \ce{CO_2} (CO2R) into higher-value products is an emerging and promising method used in the global shift towards a sustainable energy and chemical economy. Converting \ce{CO_2} into useful fuels and high-value chemicals directly removes a harmful greenhouse gas from the atmosphere. 
Renewable energy technologies, notably wind and solar, have made significant advancements in cost; in many regions of the world, these have become the cheapest available source of energy \citep{afif2022ultra, corral2021advanced}.
However, renewable energy sources are known to suffer from unpredictable intermittency that prevents scale-up to grid-scale levels. CO2R offers a solution to address this challenge by enabling the conversion of excess renewable energy into versatile storage and transportation forms \citep{corral2022bridging}
As an added benefit, the ability to power the electrochemical CO2R process with a renewable energy source opens a possible sustainable pathway, where chemical processing and energy storage can be achieved while mitigating additional greenhouse gas emissions. 

While \ce{CO_2} electrolyzers have made significant strides as a technology \citep{nitopi2019progress, stephens20222022}, further progress is needed to achieve performance at commercial-scale and cost-effective levels \citep{wakerley2022gas, park2021towards}. Copper-based catalysts are of interest due to their unique ability to produce high-value hydrocarbons, aldehydes, and alcohols in substantial amounts (Faradaic efficiencies) \citep{nitopi2019progress, kuhl2012new}. However, electrolyzers using copper catalysts suffer from high overpotential (low energy efficiencies) at desired current densities and limited selectivity towards the desired high-value products \citep{nitopi2019progress}. Issues known to contribute to this limited performance include local reaction environment features that promote hydrogen evolution reaction (HER) \citep{burdyny2019co} and other unwanted side reactions, limited \ce{CO_2} availability at the catalyst surface \citep{duarte2019electrochemical}, and ineffective management of intermediate species transport \citep{zhang2022highly}. 
Further investigation into these (and more) physical phenomena is needed towards designing high-performance CO2R electrolyzers. To this end, computational simulations have emerged as a powerful tool for investigating the high complexity of the underlying physicochemical processes and the wide landscape of reactor designs found in \ce{CO_2} electrolysis.
Simulation capabilities that use continuum modeling focus on the larger-scale behavior while representing the atomistic-scale information with an appropriate low-order description (such as using Butler-Volmer equations to model the kinetics). 
Such continuum modeling simulations have emerged as a set of powerful tools in recent years and have demonstrated the ability to provide insight into relevant electrolyzer factors, including the aforementioned issues connected with performance degradation. Notably, previous studies have provided information on the interactions between mass/ion transport, fluid flow, reaction kinetics \citep{bui2022continuum, moore2021elucidating}, key design parameters such as porosity and catalyst loading \citep{weng2018modeling}, and structural details of the electrical double layer \citep{bohra2019modeling}. Understanding these factors is important for describing selectivity and activity relationships in \ce{CO_2} electrolysis, particularly in custom reactor design architectures.


In electrolyzers, the previously mentioned ineffective management of intermediate species transport presents itself through a range of symptoms, including excess intermediate species that are advected away before being utilized \citep{zhang2022highly, wei2023nanoscale} and as diffusion/mixing-caused losses for large spacings between electrodes \citep{gurudayal2019sequential}. Specifically for CO2R reaction pathways on copper, carbon monoxide (\ce{CO}) is the crucial intermediate species for the production of further reduced high-value products (i.e., requiring $>2$\ce{e^-} transfers)\citep{nitopi2019progress}.
To address these transport limitations for intermediate species, designs that utilize multiple catalysts in proximity to synergistically leverage the reaction strengths of each catalyst material have attracted recent research attention \citep{lum2018sequential, zhang2022tandem, zhu2021tandem, lin2022vapor}. In such designs, one catalyst is effective at converting a given reactant species to an intermediate species, and a second catalyst is effective at converting the intermediate species to the desired end product(s). With the catalysts placed in proximity, the intermediate species is transported directly from the first to the second catalyst without the need for intermediate isolation or other specialized processing steps. The reduced travel distance of the intermediate species, compared to that of traditional reactor designs, results in increased residence time and availability of the intermediate species near the second catalyst (i.e., \ce{CO} near the \ce{Cu} catalyst for CO2R \citep{zhang2022highly}), allowing for increased utilization that results in higher production and selectivity towards desired products than achieved by a single catalyst setup.
This described approach to catalyst patterning is known as tandem catalysis, which is how it shall be referred to in this paper. Note that in other published studies, the approach has also been referred to as cascade, sequential, domino, spillover, or sequential cascade catalysis.
For CO2R, the product distribution for polycrystalline metal electrode catalysts was first comprehensively quantified in studies by Hori et al. \citep{hori1985production, hori1986production, hori1988enhanced, hori1989formation, hori1994electrocatalytic} using an electrolysis cell with \ce{CO_2}-saturated $0.1$ M \ce{KHCO_3} electrolyte at $5.0$ mA/cm$^2$. These studies established that \ce{Au}, \ce{Ag}, \ce{Zn}, \ce{Pe}, and \ce{Ga} catalysts primarily produce \ce{CO}; notably, \ce{Au} and \ce{Ag} perform especially well, boasting Faradaic efficiency values towards \ce{CO} of $>80$\%. The majority of the other tested metal catalysts are categorized as either mainly producing formate (\ce{HCOO^-}) or mainly reducing water to hydrogen (\ce{H_2}). Distinctively, \ce{Cu} is unique in its ability to produce high-value products, including \ce{C_{2+}} products, at appreciable Faradaic efficiencies.
Given the rationale and goals of tandem catalysis described previously, in CO2R it is thus sensible to choose \ce{Au} or \ce{Ag} as the first catalyst (to perform \ce{CO_2 -> CO}) and \ce{Cu} as the second catalyst (to perform \ce{CO -> } desired high-value final products including ethylene and ethanol). Therefore, this combination of \ce{Au}/\ce{Ag} + \ce{Cu} catalysts has been utilized in numerous tandem catalyst studies, covering both experimental and computational approaches, in the CO2R literature.
The catalysts setups of these studies span a range of different designs, including interspersed spheres/particles \citep{chen2020cu, wang2018co2}, alternating planar sections \citep{gurudayal2019sequential, lum2018sequential}, dots on planar sheets \citep{lum2018sequential}, particles interspersed within a bed of needles \citep{wei2023nanoscale}, nanoparticles on a reactive flat substrate \citep{morales2018improved}, and segmented catalyst layers of a gas diffusion electrode (GDE) \citep{zhang2022highly}, among others. These studies and the overall body of work have provided key preliminary insight into the electrochemical behaviors and the effective design principles for tandem catalysis applied to CO2R electrolyzers.


Thus far, CO2R tandem catalysis studies, including the examples listed above, have been primarily limited to \emph{forward analysis}: a specific setup or patterning of catalysts is proposed and evaluated in an electrolyzer setting (either experimentally or through simulations), but the resulting insights are not directly used to update or improve the design. In contrast, \emph{design optimization} methods \citep{martins2021engineering, tortorelli2010solid} offer a systematic framework for exploring the electrolyzer design space and improving performance. By exploiting the sensitivity of the performance metric to the catalyst patterning, the design can be iteratively refined and the problem re-solved until reaching an optimal configuration.
A natural formulation of this approach is as a PDE-constrained optimization problem, in which the governing transport and reaction equations act as constraints while the catalyst patterning serves as the design variable. Such methods have been applied in a number of applications, including shape/topology optimization for aerospace structural design \citep{giles2000introduction}, multi-body robotics systems \citep{nachbagauer2015use}, and even electrochemical systems \citep{roy2022topology, yaji2018topology, alizadeh2024recent, lin2022topology}. However, the use of design optimization in electrochemical systems remains relatively nascent, presenting a significant opportunity to leverage these techniques to guide the spatial arrangement of catalysts for enhanced CO2R performance.


The study presented in this paper combines the capabilities of 1) computational simulations with continuum transport modeling and 2) design optimization, within a simplified case study of tandem catalysis for CO2R. Our setup involves a two-dimensional (2D) flow cell plane reactor setup with \ce{Ag} and \ce{Cu} electrode catalyst surfaces. The procedures, performance metrics, and concentration flow fields are presented for designs that optimize the output of desired products across various relevant input parameters: electrode applied voltage, electrolyte flow rate, and degree of patterning (the total number of \ce{Ag}/\ce{Cu} catalyst sections). It is demonstrated that the optimal patterning design strongly depends on each of these input parameters. Noticeable improvements in the current density towards desired products such as ethylene can be achieved, especially at more negative applied voltages and by using high patterning frequencies (i.e., a large number of \ce{Ag}/\ce{Cu} sections).

The results from this study provide general insight into optimal reactor design and concentration field behavior for tandem catalyst setups. However, realistic, industrially-relevant CO2R designs include complicated and sophisticated components, such as gas diffusion electrodes (GDEs), whose effects are not fully captured by our current setup. Nevertheless, this study is intended to provide a proof-of-concept and first step towards optimization of more realistic electrolyzer geometries.

The remainder of the paper is organized as follows. \Cref{sec:Methods} presents the formulation of the problem as well as the computational details of the governing equations (for the chemistry, species transport, and fluid) and of the optimization methodology. \Cref{sec:Results_and_discussion} shows the results of the study and associated discussion. Finally, \cref{sec:Conclusions} provides conclusions and thoughts regarding future directions of work.

\section{System description and methodology}\label{sec:Methods}

\subsection{Domain setup and problem formulation}\label{sec:Domain_setup_and_problem_formulation}

\begin{figure}[!h]
    \centering
    \includegraphics[width=\textwidth]{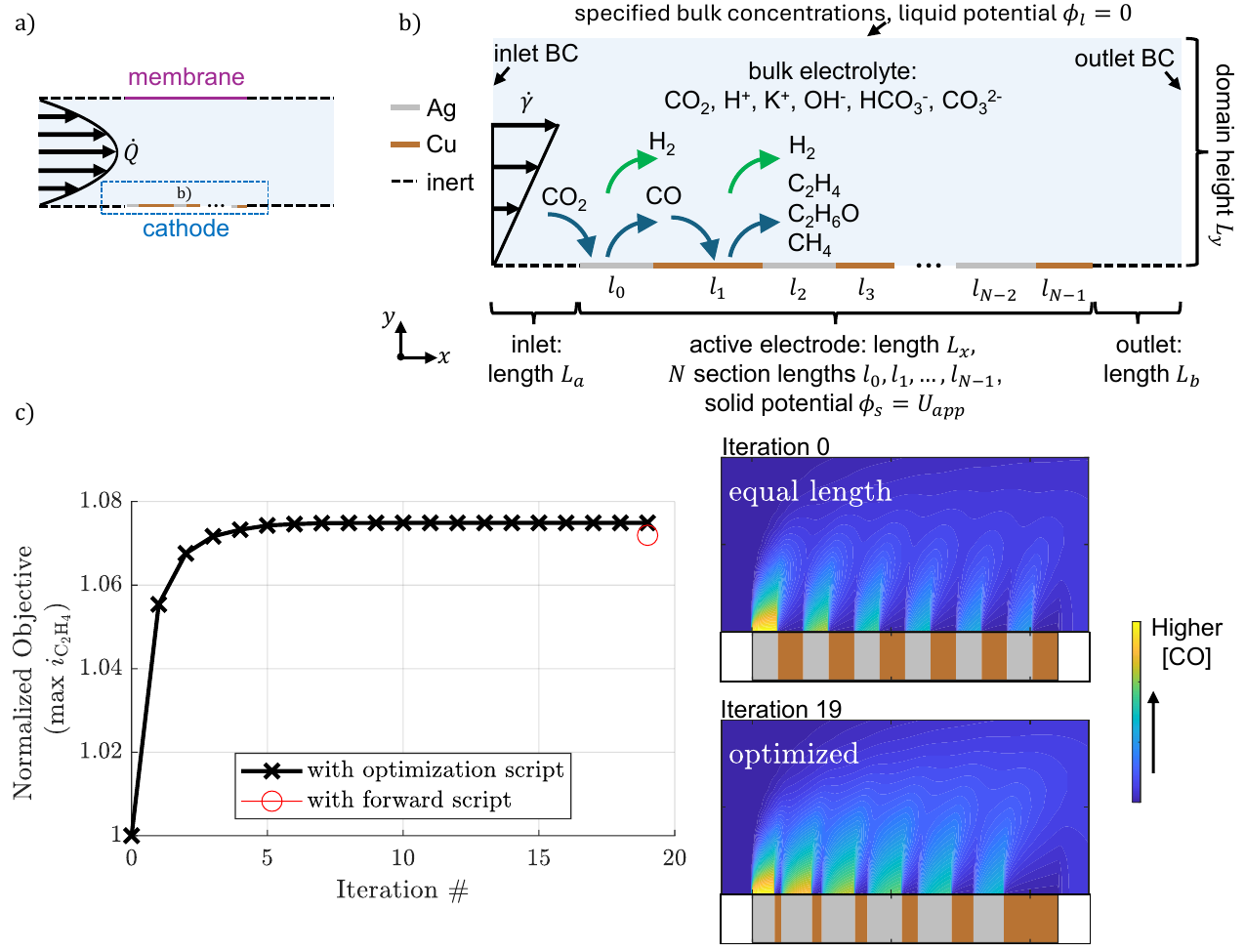}
    \caption{\doublespacing a) Schematic of a general flow reactor configuration, where $\dot{Q}$ is the volumetric flow rate flowing between two flat plates. We focus on the mass transfer phenomena near the cathode, as highlighted in panel b. b) Schematic of computational simulation setup, with a shear flow of aqueous electrolyte solution flowing over the cathode portion and the inlet/outlet regions. c) Plot of the normalized objective function -- the maximum ethylene current density, calculated per \cref{eq:domain_averaged_i} -- for each optimization iteration step number, for an example optimization case (parameters: $N = 12$, flow rate = $3.0$ ml/min, $U_{app} = -1.7$ V vs. SHE). The \ce{Ag}/\ce{Cu} section lengths and example \ce{CO} concentration fields for the (initial) equal length pattern at iteration $0$ and for the final optimized pattern at iteration $19$ are shown.}
    \label{fig:schematic_setup_and_opt}
\end{figure}

\Cref{fig:schematic_setup_and_opt} provides a schematic overview of the setup used in this work. \Cref{fig:schematic_setup_and_opt}a shows a diagram of an example flow reactor half cell. Liquid electrolyte is flowed in between the cathode and the membrane with a parabolic velocity profile (for plane Poiseuille flow) at a prescribed volumetric flow rate.

Local features such as the local \ce{CO_2} and \ce{CO} concentrations, pH, and boundary layer effects are critical ingredients that determine the yields of the desired products in the reactor system; thus, the computational setup was designed with the focused study of such features in mind. \Cref{fig:schematic_setup_and_opt}b shows the computational simulation domain used in this work, which comprises the zoomed-in region near the cathode of the half cell in \cref{fig:schematic_setup_and_opt}a. The domain is modeled as a two-dimensional (2D) region. The fluid behavior near the electrolyte/electrode interface is well-approximated as a shear flow with shear rate $\dot{\gamma}$, as is discussed in further detail in \cref{sec:Fluid_flow_modeling}.

In the bulk region above the interface (for $0<y<L_y$), an aqueous electrolyte solution flows in at the inlet, consisting of $500$ mM \ce{KHCO_3} and $34$ mM \ce{CO_2}.  A number of species -- \ce{CO_2}, \ce{H^+}, \ce{OH^-} \ce{HCO_3^-}, and \ce{CO_3^{2-}} -- participate in the bicarbonate bulk reactions and are also transported throughout the bulk region. \ce{K^+} does not participate in the bulk reactions but is still present in the bulk and so must be treated as a transported species. Details of the bulk electrolyte reactions and transport governing equations are provided in \cref{sec:Bulk_reactions_and_transport}.

The solid surface of the computational domain (at $y = 0$) consists of an active electrode region with length $L_x$, which is located in between two inert (non-reactive) portions that serve to minimize the computational boundary effects and to better capture the large fluxes at the beginning and end of the electrode. The inert portions consist of a leading inlet with length $L_a$ and a trailing outlet with length $L_b$. For this study, $L_x = 1.1$ cm, and $L_a = L_b = L_x/10$. As discussed in \cref{sec:Fluid_flow_modeling}, the example 3D flow channel that this study is based on has a length of $2.2$ cm. To yield a computationally manageable case, the length was reduced by $50$\% for the value of $L_x$. For a given total number of sections $N$ (an input parameter of the system), the active electrode region is comprised of $N$ \ce{Ag}/\ce{Cu} sections, with the patterning determined by the section lengths $l_j$ for $j = 0, 1, ..., N-1$. \Cref{fig:schematic_setup_and_opt}b also schematically shows the reactions included in the setup. The \ce{Ag} section features a \ce{CO_2 -> CO} surface reaction, and the \ce{Cu} section surface chemistry include \ce{CO -> C_2H_4}, \ce{CO -> C_2H_6O}, and \ce{CO -> CH_4} surface reactions. Both \ce{Ag} and \ce{Cu} sections also include the hydrogen evolution reaction to produce \ce{H_2}. Details of the electrochemical surface reaction model formulation are provided in \cref{sec:Surface_Ag_catalyst_chemistry} and \cref{sec:Surface_Cu_catalyst_chemistry}. We stress that \ce{CO} is not included in the input electrolyte stream, and so for this study's setup, any \ce{CO} in the system must be generated as a product by the \ce{Ag} sections and is then used as a reactant by the \ce{Cu} sections. These species that have been newly introduced by the surface reactions do not participate in the bulk reactions, but their transport within the bulk region is still modeled, as is discussed in \cref{sec:Bulk_reactions_and_transport}. To make use of the cascade/tandem nature of the electrode surface reactions, the patterns in this study consist of repeating pairs of \ce{Ag} and \ce{Cu} sections: \ce{Ag}, \ce{Cu}, \ce{Ag}, \ce{Cu}, ..., \ce{Ag}, \ce{Cu}; with this choice, only even values of $N$ are considered.

The boundary conditions for the system are also schematically listed in \cref{fig:schematic_setup_and_opt}b; these are explained in detail in \cref{sec:Bulk_reactions_and_transport}, \cref{sec:Surface_Ag_catalyst_chemistry}, and \cref{sec:Surface_Cu_catalyst_chemistry}.

\Cref{fig:schematic_setup_and_opt}c shows an optimization case at an example set of conditions. The objective function, chosen to be the maximum ethylene current density as calculated per \cref{eq:domain_averaged_i}, is plotted and is shown to be maximized as the optimization algorithm modifies the section lengths $l_j$ with each successive iteration step until convergence. The beginning and end patterns, with the corresponding example \ce{CO} concentration fields, are also shown. The initial pattern configuration is set as equal section lengths for all sections (\ce{Ag} and \ce{Cu}), while the final optimized pattern configuration (at iteration step $19$) is a nontrivial configuration with varied section lengths. Analysis and discussion of the equal length case versus optimized case, for this set of conditions and for other conditions, are provided in \cref{sec:Results_and_discussion}. Details of the optimization methodology are described in \cref{sec:Optimization_methodology}. Note that in the objective vs. iteration \# plot, the value of the normalized objective obtained from the optimization script is plotted for each iteration step. Additionally, the normalized objective value obtained from the forward script (which is treated as the true objective function value) is also plotted for the final iteration step and has a slightly different value. This detail is a consequence of the optimization procedures and is also explained in \cref{sec:Optimization_methodology}.

\subsection{Governing equations}\label{sec:Governing_equations}

\subsubsection{Bulk (bicarbonate electrolyte) reactions and transport }\label{sec:Bulk_reactions_and_transport}

The following set of governing reactions describes the buffer chemistry occurring in the bulk bicarbonate electrolyte \cite{schulz2006determination}:
\begin{subequations}
    \label{eq:bulk_rxns}
    \begin{align}
        \ce{CO_2 + OH^-  <=>[k_{1f}][k_{1r}] HCO_3^- }, \\
        \ce{CO_2 + H_2O  <=>[k_{2f}][k_{2r}] HCO_3^- + H^+}, \\
        \ce{CO_3^{2-} + H_2O  <=>[k_{3f}][k_{3r}] HCO_3^- + OH^-}, \\
        \ce{CO_3^{2-} + H^+ <=>[k_{4f}][k_{4r}] HCO_3^-}, \\
        \ce{OH^- + H^+ <=>[k_{5f}][k_{5r}] H_2O}.  
    \end{align}
\end{subequations}
$k_{jf}$ and $k_{jr}$ are the forward and reverse reaction rates, respectively, for reaction $j$. The steady equilibrium Nernst-Planck (NP) equations describe the species transport. For the $N_{\text{species}}$ total species, the equation for each species $k$ is written as
\begin{equation}\label{eq:partial_c_all}
    0 = - \nabla \cdot \mathbf{J}_{ k } + \sum_j R_{j,k},
\end{equation}
where the reaction term $R_{j,k}$ for reaction $j$ and species $k$ is defined as
\begin{equation}
    R_{j,k} = s_{k,j} c_{ref} \left( k_{jf} \prod_{m, s_{m,j} < 0} a_m^{-s_{m,j}} - k_{jr} \prod_{m, s_{m,j} > 0} a_m ^{s_{m,j}} \right),
\end{equation}
where $s_{k,j}$ is the stoichiometric coefficient and $a_{m} = c_m / c_{ref}$ is the activity of species $m$ given concentration $c_m$ and reference concentration $c_{ref} = 1$M. $\mathbf{J}_k$ is the flux of species $k$, defined as
\begin{equation}\label{eq:J_i}
    \mathbf{J}_k = -D_k \nabla c_k - z_k u_k F c_k \nabla \phi_l + \mathbf{v} c_k.
\end{equation}
For species $k$, $c_k$ represents the concentration, $D_k$ is the diffusivity, $z_k$ is the valency, and $u_k$ is the ionic mobility, given from the Nernst-Einstein relation and written as 
\begin{equation}
    u_k = \frac{D_k}{R T},
\end{equation}
where $R$ is the universal gas constant and $T$ is the thermodynamic temperature. $F$ is Faraday's constant, $\phi_l$ is the fluid (electrolyte) potential, and $\mathbf{v}$ is the fluid (electrolyte) flow velocity. $\mathbf{v}$ is modeled as a simple shear flow with shear rate $\dot{\gamma}$, 
\begin{align}\label{eq:fluid_velocity}
    \mathbf{v} &= \begin{bmatrix}
       \dot{\gamma} y \\
       0
     \end{bmatrix},
\end{align}
as discussed in more detail in \cref{sec:Fluid_flow_modeling}. Section S1 of the Supplementary Information document provides the tabulated diffusivity values (in table S1) and reaction rate constant values (in table S2).

These $N_{\text{species}}$ transport equations are used to solve for $N_{\text{species}}+1$ unknown variables ($N_{\text{species}}$ concentrations $c_1, c_2, ..., c_{N_{\text{species}}}$ in addition to the fluid potential $\phi_l$). The additional equation needed to complete the system is the electroneutrality approximation:
\begin{equation}
    \sum_k z_k c_k = 0.
\end{equation}

\paragraph{Boundary conditions}
The following boundary conditions are used:
\begin{enumerate}
    \item Bottom of domain ($y = 0$): catalyst surface chemistry; this provides boundary conditions for the ionic current $\mathbf{i_2}$,
    \begin{equation}
        \mathbf{i_2} \cdot \mathbf{n} = - \sum_{p} i_{\ce{p}, \text{local}},
    \end{equation}
    and for the mass flux $\mathbf{N}_k$ for species $k$,
    \begin{equation}
        \mathbf{N}_k \cdot \mathbf{n} = - \sum_{p} \frac{s_{k,p} i_{\ce{p}, \text{local}}}{n_p F},
    \end{equation}
    where $n$ represents the number of electrons transferred and $p$ is the index for each charge-transfer reaction in the system. 
    The Tafel expressions for $i_{\ce{p}, \text{local}}$ are provided in \cref{sec:Surface_Ag_catalyst_chemistry} and \cref{sec:Surface_Cu_catalyst_chemistry}. For the inert inlet and outlet regions, $\mathbf{i_2} \cdot \mathbf{n} = \mathbf{N}_k \cdot \mathbf{n} = 0$.
    \item Top of domain ($y = L_y$):
    \begin{enumerate}
        \item $\phi_l = 0$
        \item $c_{\ce{CO_2} } = 34$ mM (from Henry's constant at a pressure of 1 atmosphere)
        \item $c_{\ce{K^+}} = 500$ mM in this study; then, $c_{\ce{OH^-} }, c_{\ce{HCO_3^-}}, c_{\ce{CO_3^{2-}}}, c_{\ce{H^+}}$ are determined at chemical equilibrium from the bulk reactions provided in \cref{eq:bulk_rxns}
        \item $c_{\ce{CO}} = c_{\ce{H_2}} = c_{\ce{CH_4}} = c_{\ce{C_2H_4}} = c_{\ce{C_2H_6O}} = 0$
    \end{enumerate}
    \item Inlet: $\mathbf{J}_k \cdot \mathbf{n} = c_{k, bulk} \mathbf{v} \cdot \mathbf{n}$
    \item Outlet: $\mathbf{J}_k \cdot \mathbf{n} = c_{k} \mathbf{v} \cdot \mathbf{n}$
\end{enumerate}
Here, $\mathbf{J}_k$ is the flux and $c_{k, bulk}$ is the bulk concentration of species $k$, and $\mathbf{n}$ is the unit outward normal vector.

\subsubsection{Surface (\texorpdfstring{\ce{Ag}}{Ag} catalyst) chemistry}\label{sec:Surface_Ag_catalyst_chemistry}
The surface chemistry on the \ce{Ag} catalytic surface is governed by the following reactions:
\begin{align}\label{eq:surface_rxn_1_Ag}
    \ce{CO_2 + H_2O + 2e^- -> 2 OH^- + CO},
\end{align}
\begin{align}\label{eq:surface_rxn_2_Ag}
    \ce{ 2 H_2O + 2e^- -> H_2 + 2 OH^-}.
\end{align}

The Tafel expressions \citep{corpus2023coupling} to model this surface chemistry in \cref{eq:surface_rxn_1_Ag} and \cref{eq:surface_rxn_2_Ag} are given as the following:
\begin{equation}\label{eq:i_CO_Tafel}
\begin{split}
    i_{ \ce{CO}, \text{local} } = \left( \frac{a_{ \ce{ CO_2} }}{a_{ \ce{CO_2} }^{bulk}} \right)^{-\gamma_{ \ce{CO_2},\ce{CO} }} \left( \frac{a_{ \ce{OH^-} }}{a_{ \ce{OH^-} }^{bulk}} \right)^{-\gamma_{ \ce{OH^-},\ce{CO} }} i_{0,\ce{CO}} \left( - \exp\left[ - \frac{\alpha_{c,\ce{CO}} F}{RT} \eta_{s,\ce{CO}} \right] \right),
\end{split}
\end{equation}

\begin{equation}\label{eq:i_H2_Tafel}
\begin{split}
    i_{\ce{H_2}, \text{local}} = \left( \frac{a_{ \ce{OH^-} }}{a_{ \ce{OH^-} }^{bulk}} \right)^{-\gamma_{ \ce{OH^-},\ce{H_2} }} i_{0, \ce{H_2} } \left( - \exp\left[ - \frac{\alpha_{c, \ce{H_2} } F}{RT} \eta_{s, \ce{H_2} } \right] \right).
\end{split}
\end{equation}
The parameters (definitions and values) and validation of the implementation are provided in sections S3 and S4 of the Supplementary Information document.

\subsubsection{Surface (\texorpdfstring{\ce{Cu}}{Cu} catalyst) chemistry}\label{sec:Surface_Cu_catalyst_chemistry}

The surface chemistry on the \ce{Cu} catalytic surface is governed by the following reactions:
\begin{align}\label{eq:surface_rxn_1_Cu} 
    \ce{2CO + 6H_2O + 8e^- -> C_2H_4  + 8 OH^-},
\end{align}
\begin{align}\label{eq:surface_rxn_2_Cu} 
    \ce{2CO + 7H_2O + 8e^- -> C_2H_6O  + 8OH^-},
\end{align}
\begin{align}\label{eq:surface_rxn_3_Cu} 
    \ce{CO + 5H_2O + 6e^- -> CH_4 + 6OH^-},
\end{align}
\begin{align}\label{eq:surface_rxn_4_Cu}
   \ce{2 H_2O + 2e^- -> 2 OH^- + H_2}. 
\end{align}

These Tafel expression formulations \citep{li2021electrokinetic} to model this surface chemistry in equations \cref{eq:surface_rxn_1_Cu} through \cref{eq:surface_rxn_4_Cu} are given as the following:
\begin{equation}\label{eq:i_C2H4_Tafel}
\begin{split}
    i_{ \ce{C_2H_4}, \text{local} } = \left( \frac{c_{\ce{CO}}}{c_{ref}} \right) i_{0, \ce{C_2H_4} } 
    \left( - \exp\left[ - \frac{\alpha_{c, \ce{C_2H_4} } F}{RT} \eta_{s, \ce{C_2H_4} } \right] \right),
\end{split}
\end{equation}

\begin{equation}\label{eq:i_C2H6O_Tafel}
\begin{split}
    i_{ \ce{C_2H_6O}, \text{local} } = \left( \frac{c_{ \ce{CO} }}{c_{ref}} \right) i_{0, \ce{C_2H_6O} } \left( - \exp\left[ - \frac{\alpha_{c, \ce{C_2H_6O} } F}{RT} \eta_{s, \ce{C_2H_6O} } \right] \right),
\end{split}
\end{equation}

\begin{equation}\label{eq:i_CH4_Tafel}
\begin{split}
    i_{ \ce{CH_4}, \text{local} } = \left( \frac{c_{ \ce{CO} }}{c_{ref}} \right) i_{0, \ce{CH_4} } \left( - \exp\left[ - \frac{\alpha_{c, \ce{CH_4} } F}{RT} \eta_{s, \ce{CH_4} } \right] \right),
\end{split}
\end{equation}

\begin{equation}\label{eq:i_H2_Tafel_Cu}
\begin{split}
    i_{ \ce{H_2}, \text{local} } = i_{0,\ce{H_2}} \left( - \exp\left[ - \frac{\alpha_{c,\ce{H_2}} F}{RT} \eta_{s,\ce{H_2}} \right] \right).
\end{split}
\end{equation}

The parameters (definitions and values) and validation of the implementation are provided in sections S5 and S6 of the Supplementary Information document.

An assumption for simplicity utilized in this study is that the \ce{CO_2 -> CO} capability on \ce{Cu} is omitted. The justification for this choice arises from the following arguments:
\begin{enumerate}
    \item Except at less negative applied voltages ($U_{app} \gtrsim -1.2$ V vs. SHE), the \ce{CO} current density values on \ce{Ag} (using the model parameters in this study \citep{corpus2023coupling}) are significantly larger than published values of the \ce{CO} net current density values on \ce{Cu} \citep{kuhl2012new}. More quantitatively, the reported net \ce{Cu} $i_{ \ce{CO}, \text{local} }$ values do not exceed $\simeq 0.2$ mA/cm$^2$, while the \ce{Ag} $i_{ \ce{CO}, \text{local} }$ modeled values in this study's voltage range span $\simeq 1.5$ to $5$ mA/cm$^2$.
    \item As observed in the CO2R on \ce{Cu} literature \citep{kuhl2012new}, the current densities of \ce{H_2}, \ce{C_2H_4}, \ce{C_2H_6O} are larger than the \ce{CO} net current density in the range of applied voltages considered in this study. \ce{CH_4} values are comparable to the \ce{CO} net values at less negative applied voltages but quickly become much larger moving towards more negative applied voltages. More quantitatively, at more negative applied voltages, the \ce{H_2}, \ce{C_2H_4}, and \ce{CH_4} current density values are at least $10 \times$ as large and the \ce{C_2H_6O} current density values are $\simeq 5-10 \times$ as large as the net \ce{CO} current density values.
    \item It is difficult, experimentally, to obtain accurate \ce{CO_2 -> CO} kinetics data on \ce{Cu} that is not compromised by \ce{CO_2} mass transport limitations. It is thus a challenge to obtain reliable Tafel parameters to be used for computational modeling. Previous modeling studies have typically turned to simplified model representations of the \ce{Cu} CO2R kinetics, such as assuming a single-step $6$-electron electrochemical reaction \citep{corral2022bridging} which focuses on capturing trends from modifications to the input parameters. 
\end{enumerate} 

If the \ce{CO_2 -> CO} kinetics on \ce{Cu} capability were to be included and considered, it is anticipated that the optimized designs would favor longer \ce{Cu} section lengths than are obtained with this study's setup; however, the qualitative takeaway messages from the design trends should still hold.

\subsection{Fluid flow setup}\label{sec:Fluid_flow_modeling}

The fluid flow velocity $\mathbf{v}$ is modeled as a shear flow with shear rate $\dot{\gamma}$, per \cref{eq:fluid_velocity}. The value of $\dot{\gamma}$ is obtained using the P\'eclet number
\begin{equation}\label{eq:Peclet_number}
    Pe = \frac{\dot{\gamma} L_x^2}{D},
\end{equation}
where $D$ is a representative diffusion coefficient value. $Pe$ quantifies the advective transport strength by the flow relative to the diffusive transport strength. Given an experimental flow rate, details of determining $\dot{\gamma}$, which then yields $Pe$ and $\mathbf{v}$ for the computational setup, are provided in section S2 of the Supplementary Information document.

Flow rate values from $3.0$ to $30.0$ ml/min represent a reasonable range of experimental flow rates. For the computational cases in this study, a value of $Pe = 7.23 \times 10^6$ (corresponding to $3.0$ ml/min) is used to represent a low flow rate, and a value of $Pe = 7.23 \times 10^7$ (corresponding to $30.0$ ml/min) is used to represent a high flow rate.

\subsection{Optimization methodology}\label{sec:Optimization_methodology}

\subsubsection{Optimization parameterization and control variable definition}

For a total number of sections $N$, each section's length is characterized using a set of $N$ variables $l_j$, for $j = 0$ to $j = N - 1$. The optimization algorithm directly acts upon a set of $N-1$ control variables $\rho_j$, for $j = 0$ to $j = N-2$. Each length $l_j$ is determined from the control variables $\rho_j$ by the following procedure:
\begin{enumerate}\label{list:rho_to_l_procedure}
    \item The first section length: $l_0 = \rho_0 \rho_1  ... \rho_{N-2}$.
    \item The intermediate section lengths (for $j = 1$ to $j = N-2$): $l_j = \rho_0 \rho_1 ... \rho_{N-(j+2)} (1 - \rho_{N - (j+1)})$. 
    \item The last section length: $l_{N-1} = 1 - \rho_0$.
\end{enumerate}
Note that for $N = 2$, there are no intermediate sections, and so $l_0 = \rho_0$ and $l_1 = l_{N-1} = 1 - \rho_0$.

This procedure has been previously used in the design optimization literature \citep{sigmund1997design, watts2016n}, where it is referred to as the volume fraction relationships for $N$ materials.

For regularization, section boundaries for each section are specified with a smoothed indicator function:
\begin{equation}\label{eq:tanh_function}
    \chi_j (x) = 0.5 \left( \tanh(S(x - b_{\text{left},j}))  - \tanh(S(x - b_{\text{right},j})\right), 
\end{equation}
where $S$ is a sharpening parameter, and $b_{\text{left},j}$ and $b_{\text{right},j}$ are the left and right bounds, respectively, of each section $j$. As $S \rightarrow \infty$, $\chi_j (x)$ approaches an ideal indicator function (i.e., with sharp section definitions rather than smooth transition regions). While well-defined sections most accurately represent the intended catalyst surface boundary condition, using values of $S$ that are too large results in undefined derivative values at the section boundaries, causing the optimization algorithm to fail to converge when computing the gradients needed for the adjoints. As a compromise, a sufficiently large value ($S > 10 / l_j$) is used, allowing the sections to be distinctly defined while minimizing the extent and impact of intermediate values ($\chi_j$ values between $0$ and $1$).

\subsubsection{Optimization procedure}\label{sec:Opt_Procedure}

The mathematical optimization problem is formulated as
\begin{equation}\label{eq:opt_problem}
\begin{aligned}
    \max_{\{\rho_j\}_{j=0}^{N-2}} \quad & \frac{1}{L_x} \int_0^{L_x} i_{\ce{C_2H_4},\mathrm{local}} \, \mathrm{d} x \\
    \text{s.t.} \quad & \mathcal{F}(\mathbf{u}) = 0, \\
                      & 0 \leq \rho_j \leq L_x, \text{ for }j=0,\dots,N-2,
\end{aligned}
\end{equation}
where $\mathcal{F}(\mathbf{u})$ represents the PDE system (forward problem) for the reactor, as presented in \cref{sec:Governing_equations}, for a given catalyst patterning set by $\rho_j$. The specified constraints on $\rho_j$ ensure that the total catalyst length $L_x$ is preserved and that the sections do not overlap. A high-level depiction of how the optimization procedure interacts with the reactor simulation model is provided in \cref{fig:schematic_setup_and_opt}. The values of the system input parameters (flow rate, target applied voltage $U_{max}$, $N$) are specified, and the specific objective function is chosen as the goal of the optimization procedure. In this study, maximizing the spatial average of the local current density towards ethylene is chosen as the objective function.

If the target applied voltage value $U_{max}$ is sufficiently negative, then the flow reactor solver does not converge using the initial guess. In this case, a continuation loop is used to iteratively solve the system starting from a less negative $U_{app}$ value and then gradually stepping towards $U_{max}$ using an increment value $\Delta U$. For this study, $\Delta U = 0.01$ V for $U_{app} >= -1.55$ V vs. SHE, and $\Delta U = 0.005$ V for $U_{app} < -1.55$ V vs. SHE.

Each design iteration is performed as follows:
\begin{enumerate}
    \item while $U_{app} > U_{max}$:
    \begin{enumerate}
            \item solve governing system with strong regularization (smaller $S$ in \cref{eq:tanh_function}) and without the following species: \ce{H_2}, \ce{C_2H_4}, \ce{C_2H_6O}, \ce{CH_4} (which do not affect the surface reaction performance and can thus be omitted to save cost)
        \item increment the applied voltage: $U_{app}$ \texttt{-=} $\Delta U$
    \end{enumerate}
    \item solve adjoint problem of step 1
    \item compute sensitivities
    \item obtain optimized design pattern ($l_{j,\text{opt}}$)
\end{enumerate}
This procedure is performed until convergence is satisfied, as described in \cref{sec:Computational_solver_details}.

Using the final optimized design pattern, the final forward solve is performed as follows:
\begin{enumerate}
    \item while $U_{app} > U_{max}$:
    \begin{enumerate}
        \item solve governing system with optimized design pattern ($l_{j,\text{opt}}$), weaker regularization (larger $S$ in \cref{eq:tanh_function}), and with all species as specified in \cref{sec:Bulk_reactions_and_transport}
        \item increment the applied voltage: $U_{app}$ \texttt{-=} $\Delta U$
    \end{enumerate}
    \item obtain final solution (solution fields and current densities)
\end{enumerate}

\subsection{Computational solver details}\label{sec:Computational_solver_details}

The flow and chemistry are solved using EchemFEM package \citep{roy2024echemfem, roy2023scalable}, which is a Python package that utilizes the open-source finite element method (FEM) library Firedrake \citep{ham2023firedrake}. 
For spatial discretization, standard piecewise-linear elements are used with a Streamline Upwind Petrov-Galerkin (SUPG) \citep{brooks1982streamline, elman2014element} stabilization scheme on the combined advection-migration flux~\citep{govindarajan2023coupling}. 
Additional details of the numerical implementation can be found in the references given. 

For the optimization methodology (\cref{sec:Optimization_methodology}), SciPy's minimize function \citep{2020SciPy-NMeth} is used as the minimization function, utilizing the limited-memory Broyden-Fletcher-Goldfarb-Shanno with bound constraints (L-BFGS-B) quasi-Newton method.
Gradients are obtained with the adjoint method, which is done automatically via pyadjoint~\citep{mitusch2019dolfinadjoint}.

For the convergence criteria, the default options for the tolerances (function tolerance \texttt{ftol}, gradient tolerance \texttt{gtol}, and Jacobian numerical approximation step size \texttt{eps}) in SciPy's L-BFGS-B method \citep{2020SciPy-NMeth} are used.

\section{Results and discussion}\label{sec:Results_and_discussion}

In this section, the results of the study are presented. To characterize the strengths of the surface reactions, two different applied voltage ($U_{app}$) values are considered. $U_{app} = -1.35$ V vs. the standard hydrogen electrode (SHE) represents a less negative (weaker surface reaction) case, and $U_{app} = -1.7$ V vs. SHE represents a more negative (stronger surface reaction) case.
To characterize the electrolyte shear flow strength, two different values of the P\'eclet number $Pe$ are considered, as mentioned and explained in \cref{sec:Fluid_flow_modeling}. $Pe = 7.23 \times 10^6$ (corresponding to a flow rate of $3.0$ ml/min) is used as a low flow rate case, and $Pe = 7.23 \times 10^7$ (corresponding to a flow rate of $30.0$ ml/min) is used as a high flow rate case.

\subsection{Manually-patterned \texorpdfstring{$N = 2$}{N = 2} cases}\label{sec:Manual_optimization}

\begin{figure}[!h]
    \centering
    \includegraphics[width=\textwidth]{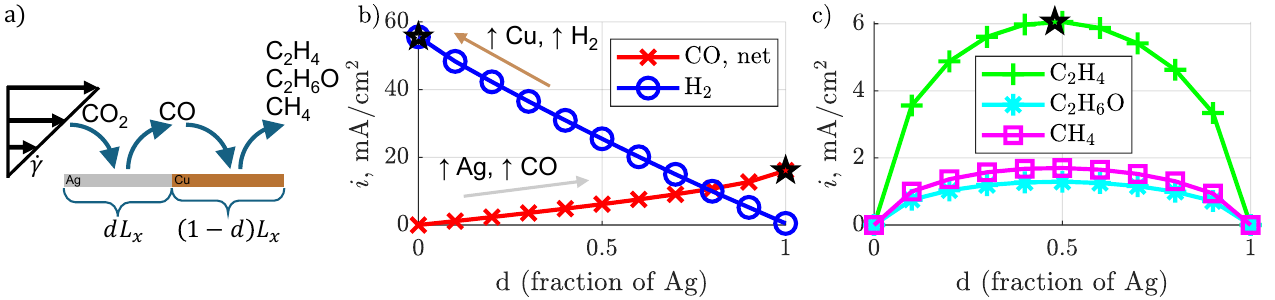}
    \caption{\doublespacing a) Schematic of the patterning configuration $N = 2$, along with the cascade reaction pathway considered: \ce{ CO_2 -> CO} for \ce{Ag}, \ce{CO -> \{C_2H_4, C_2H_6O, CH_4\} } for \ce{Cu}. b) Plots of current density for net \ce{CO} and for \ce{H_2}, shown at $11$ values of the \ce{Ag} fraction ($d = 0$ through $d = 1$ in increments of $0.1$). c) Plots of current density for \ce{C_2H_4}, \ce{C_2H_6O}, and \ce{CH_4}, in the same format as for panel b. All cases shown use the same conditions as chosen in \cref{fig:schematic_setup_and_opt}: flow rate = $3.0$ ml/min, $U_{app} = -1.7$ V vs. SHE. The maximum current density values and the corresponding value of $d$ for \ce{CO} (net), \ce{H_2}, and \ce{C_2H_4} are shown for the appropriate curve with a black star; these maximum values have also been verified using the optimization procedure as presented in \cref{sec:Optimization_methodology}.}
    \label{fig:trivial_opt_figure}
\end{figure}

As an introduction and motivation prior to showing the optimization results, \cref{fig:trivial_opt_figure} provides simulation results for cases that are set up and solved manually (using only the final forward solve procedure, skipping the optimization procedure in \cref{sec:Opt_Procedure}). All of these electrode patterns use $N = 2$ and involve one \ce{Ag} section upstream of one \ce{Cu} section, as the simplest patterning configuration available for this work's problem setup. As shown in \cref{fig:trivial_opt_figure}a, the \ce{Ag} fraction $d$ fully characterizes the patterning in this simplified configuration, creating an electrode consisting of one \ce{Ag} section with length $= d L_x$ and one \ce{Cu} section with length $= (1 - d) L_x$. The \ce{Ag} fraction $d$ is varied from $d = 0$ (for a fully \ce{Cu} electrode) to $d = 1$ (for a fully \ce{Ag} electrode), thus fully sweeping through the electrode patterning designs for $N=2$.

In \cref{fig:trivial_opt_figure}b and \cref{fig:trivial_opt_figure}c, the current density $i_k$ for each electrode product species $k$ is plotted as a function of $d$. $i_k$ is computed as a spatial average of the local current density $i_{k, local}$ (expression for each product species provided in \cref{sec:Governing_equations}) over the length of the electrode,
\begin{equation}\label{eq:domain_averaged_i}
    i_k = \frac{1}{L_x} \int_0^{L_x} i_{k,\text{local}} \, \mathrm{d} x.
\end{equation}

Note that the \ce{H_2} current density $i_{\ce{H_2}}$ has contributions from both \ce{Ag} and \ce{Cu}. An important detail is that the \ce{CO} current density shown is the net value, i.e. the quantity that would be measured in an experiment. \ce{CO} is unique among the electrode products in this study's setup in that it is both produced (on \ce{Ag} sections) and consumed (on \ce{Cu} sections); thus, only integrating the \ce{CO} current density from \cref{eq:i_CO_Tafel} would only yield contribution from the production while neglecting the consumption.  Thus, the plotted net \ce{CO} current density $i_{\ce{CO},\mathrm{net}}$ is computed from the \ce{CO} production ($i_{\ce{CO},+}$) and consumption ($i_{\ce{CO},-}$) as
\begin{align}\label{eq:CO_production_consumption}
    \begin{split}
        i_{\ce{CO},\mathrm{net}} = i_{\ce{CO},+} - i_{\ce{CO},-}, \\
        \text{with } i_{\ce{CO},+} = i_{\ce{CO}}, \qquad i_{\ce{CO},-} = \frac{i_{\ce{C_2H_4}}}{2} + \frac{i_{\ce{C_2H_6O}}}{2} + \frac{i_{\ce{CH_4}}}{3},
    \end{split}
\end{align}
in order to account for the stoichiometric coefficients for the product species and electrons that participate in the reactions. 

In \cref{fig:trivial_opt_figure}b, $i_{\ce{CO},\mathrm{net}}$ increases with \ce{Ag} content ($d \rightarrow 1$), and $i_{\ce{H_2}}$ increases with \ce{Cu} content ($d \rightarrow 0$). The optimal design is therefore trivially a pure \ce{Ag} electrode for \ce{CO} or a pure \ce{Cu} electrode for \ce{H_2}. Consequently, maximizing either \ce{CO} or \ce{H_2} current density alone does not yield a meaningful optimization problem.

In contrast, as shown in \cref{fig:trivial_opt_figure}c, $i_{\ce{C_2H_4}}$, $i_{\ce{C_2H_6O}}$, and $i_{\ce{CH_4}}$ all show a maximum value at $d \simeq 0.5$, thus showing non-trivial optimal designs.
To successfully maximize production of these products, the electrode patterning requires sufficient amounts of both \ce{Ag} (to perform the \ce{ CO_2 -> CO} conversion) and \ce{Cu} (to perform the \ce{CO -> \{C_2H_4, C_2H_6O, CH_4\} } conversion). Insufficient section lengths of either electrode material (seen for $d$ values near $0$ and near $1$) lead to inadequate contribution of one of these reaction sets in the tandem/cascade pathway, leading to overall suboptimal output. These three products -- \ce{C_2H_4}, \ce{C_2H_6O}, and \ce{CH_4} -- behave similarly to each other as a function of $d$, with only a simple scaling factor needed to approximate the current density profile of one given that of another. 
Thus, any of the three products can serve to illustrate the behavior of the \ce{Cu} hydrocarbon products as a whole. For this reason and because the hydrocarbon products represent the value-added chemicals that are sought after in CO2R, the maximum $i_{\ce{C_2H_4}}$ is chosen as a representative objective function used in the optimization simulations presented through the rest of this study.
Note that this similarity in profiles is likely a consequence of the chemistry modeling in this study. More realistic/accurate microkinetic modeling or experiments would likely show more complex relationships among the \ce{Cu} product profiles. However, because other studies also generally show a positive correlation among these major \ce{Cu} products \citep{nitopi2019progress}, the general takeaways obtained by using the \ce{C_2H_4} current density as a representative metric should still hold.

The manual optimization sweep provides preliminary insight into reactor performance and optimization objectives. While a systematic search is tractable for the single design variable case ($N=2$), it becomes computationally unfeasible as $N$ increases and the number of possible configurations grows exponentially. As demonstrated in \cref{sec:Optimal_patterns}, higher hydrocarbon current densities can be achieved at larger $N$ with optimized, often unintuitive, patterns. Nevertheless, the results here serve as a useful foundation for understanding reactor behavior prior to exploring more complex designs.

To quantify the product selectivity, fig. S4 of the Supplementary Information shows the Faradaic efficiency (FE) profile using the same flow rate and $U_{app}$ conditions as used in \cref{fig:trivial_opt_figure}. Similar qualitative behavior is observed as in the current density behavior seen in \cref{fig:trivial_opt_figure} and is discussed in section S7 of the Supplementary Information.

\subsection{Effect of number of sections \texorpdfstring{$N$}{N}, flow rate, and applied voltage \texorpdfstring{$U_{app}$}{U_{app}} on optimized pattern}\label{sec:Optimal_patterns}

\begin{figure}[!h]
    \centering
    \includegraphics[width=\textwidth]{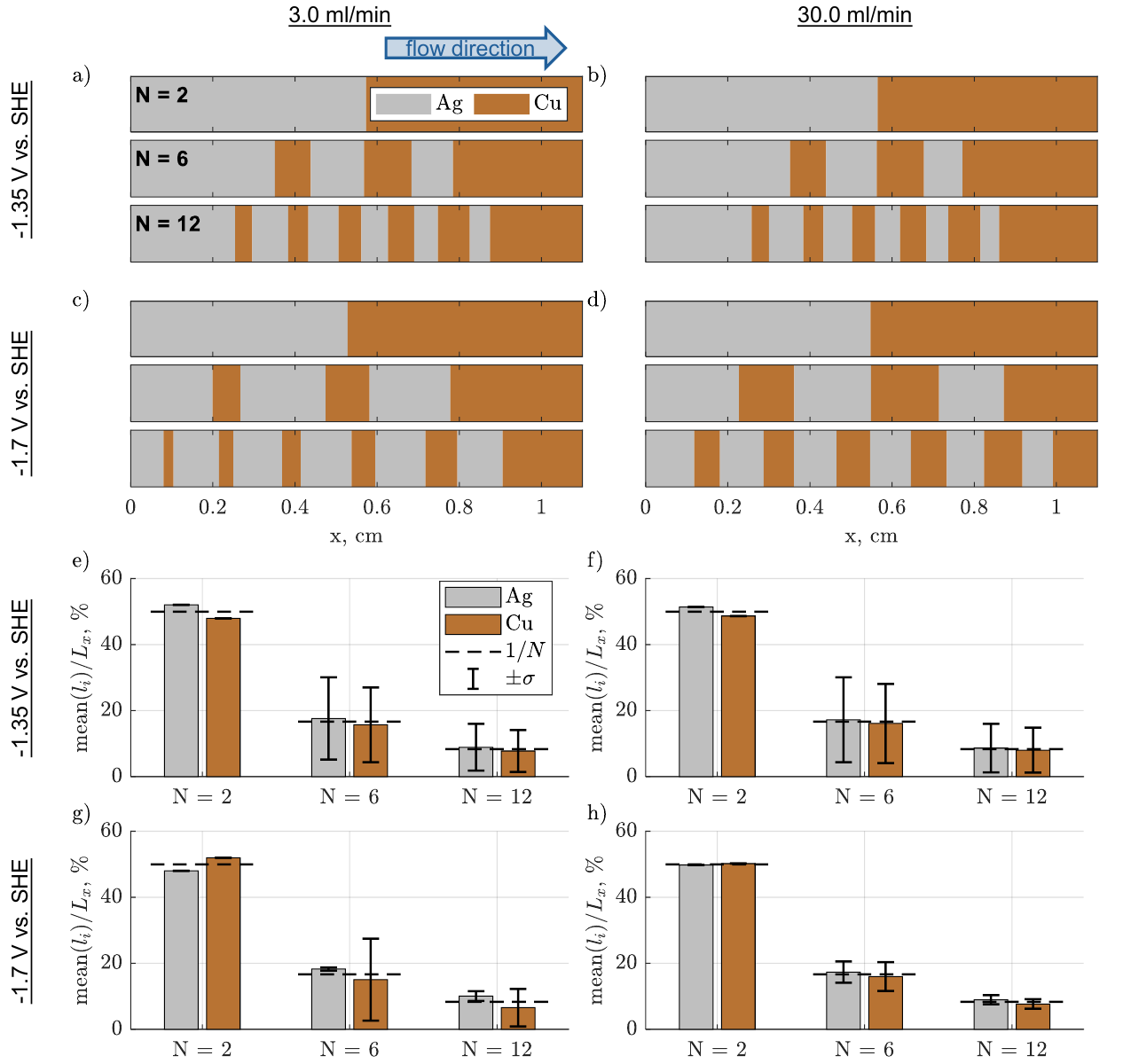}
    \caption{\doublespacing Plots of patterning and of section length statistics for optimized designs for various $N$, $U_{app},$ and flow rate values. a) - d) Optimized patterning showing section locations and length in $x$. e) - h) Mean \ce{Ag} and \ce{Cu} section lengths, normalized by electrode length $L_x$ and shown as a percentage. Horizontal dashed lines for each $N$ represent the section length for the equal length configuration ($l_j = L_x/N$ for each section $j$). Vertical error bars represent $\pm 1$ standard deviation of the \ce{Ag} and \ce{Cu} section lengths.}
    \label{fig:opt_vs_N_figure}
\end{figure}

With the insight gained from the manual optimization results in \cref{sec:Manual_optimization}, our attention now turns to the results of cases generated using the design optimization methodology from \cref{sec:Optimization_methodology}. A set of optimization case results is shown in \cref{fig:opt_vs_N_figure}. The initial configuration for all presented cases is equal section lengths ($l_j = L_x/N$ for each section $j$) with an alternating \ce{Ag}/\ce{Cu}/\ce{Ag}/\ce{Cu}/... pattern. Following the discussion in \cref{sec:Manual_optimization}, the objective function is chosen to be maximizing $i_{\ce{C_2H_4}}$ for all optimization cases shown.

The following parameter choices are used for the optimization cases shown:
\begin{enumerate}
    \item Two applied voltage $U_{app}$ conditions: The less negative value of $U_{app} = -1.35$ V vs. SHE results in a weaker surface reaction strength, and the more negative value of $U_{app} = -1.7$ V vs. SHE results in a stronger surface reaction strength.
    \item Two flow rate values are chosen: $3.0$ and $30.0$ ml/min. These represent the lower and upper bounds of an example practical experimental flow rate, as discussed in \cref{sec:Fluid_flow_modeling}.
    \item Three values of the number of sections $N$ are chosen: $\{ 2, 6, 12 \}$.
\end{enumerate}

\Cref{fig:opt_vs_N_figure}a through \cref{fig:opt_vs_N_figure}d show the patterns (section lengths and positions) for the optimized designs. For $N = 2$ designs, the optimized patterns for all conditions result in approximately $50$\% each for the \ce{Ag} and \ce{Cu} section lengths, regardless of the $U_{app}$ and flow rate conditions. For $N = 2$, it is evident that sufficient amounts of both catalysts are necessary to yield maximum \ce{C_2H_4} production; this is consistent with the findings from \cref{sec:Manual_optimization}. For the larger $N$ values of $6$ and $12$, optimized patterns arise that display more interesting features and also show distinct patterns across the different combinations of $U_{app}$ and flow rate conditions.

For the $U_{app} = -1.35$ V vs. SHE cases, the optimized patterns at larger $N$ are characterized by long lengths for the first \ce{Ag} section and the last \ce{Cu} section. The intermediate \ce{Ag} and \ce{Cu} sections in the middle of the domain alternate rapidly, with each being significantly shorter than the first and last sections. These intermediate sections show slight gradients of shortening \ce{Ag} and lengthening \ce{Cu} section length moving along the flow direction. Interestingly, at this less negative $U_{app}$ value, the optimized patterns are not appreciably different between the two different flow rate values.

For the $U_{app} = -1.7$ V vs. SHE cases, the optimized patterns at larger $N$ display different features compared to the less negative $U_{app}$ value and also show a significant dependency on flow rate. Starting with the high flow rate ($30.0$ ml/min) optimization case, the final optimized design displays only minimal deviation from the initial equal length configuration; the design also shows slight gradients of shortening \ce{Ag} and lengthening \ce{Cu} section lengths moving along the flow direction. It is observed in \cref{sec:Current_density_performance} that these more negative $U_{app}$ + high flow rate conditions correspond to the highest magnitude of \ce{C_2H_4} current density, implying that the initial equal length patterning is already high performing and close to the optimal design for these conditions. For the low flow rate ($3.0$ ml/min) optimization case, which is the most mass transport-limited case considered, the optimized patterning involves \ce{Ag} sections that are all similar in length and \ce{Cu} sections that noticeably increase in length along the flow direction.

\Cref{fig:opt_vs_N_figure}e through \cref{fig:opt_vs_N_figure}h show the mean and standard deviations of the \ce{Ag} and \ce{Cu} section lengths for the optimized case design patterns across the flow rate/$U_{app}$ conditions. Note that the $N = 2$ optimized cases involve only one \ce{Ag} section and one \ce{Cu} section, and so their standard deviation is always equal to $0$. Just as was observed in the section designs, a number of distinctive features appear in the section length mean and standard deviation values for the $U_{app} = -1.7$ V vs. SHE, flow rate $= 3.0$ ml/min conditions. Firstly, the total amount of \ce{Ag} is favored over \ce{Cu}, with the mean \ce{Ag} section being $1.53 \times$ longer than the mean \ce{Cu} for $N = 12$. Secondly, at these conditions, the standard deviation for \ce{Ag} section lengths is significantly smaller than the \ce{Cu} value; for all other conditions, the standard deviations of \ce{Ag} and \ce{Cu} section lengths are roughly equal. These attributes provide further evidence of the unique optimization behavior that occurs at the more negative $U_{app}$ and low flow rate conditions. 

\subsection{Current density performance}\label{sec:Current_density_performance}
\begin{figure}[!h]
    \centering
    \includegraphics[width=\textwidth]{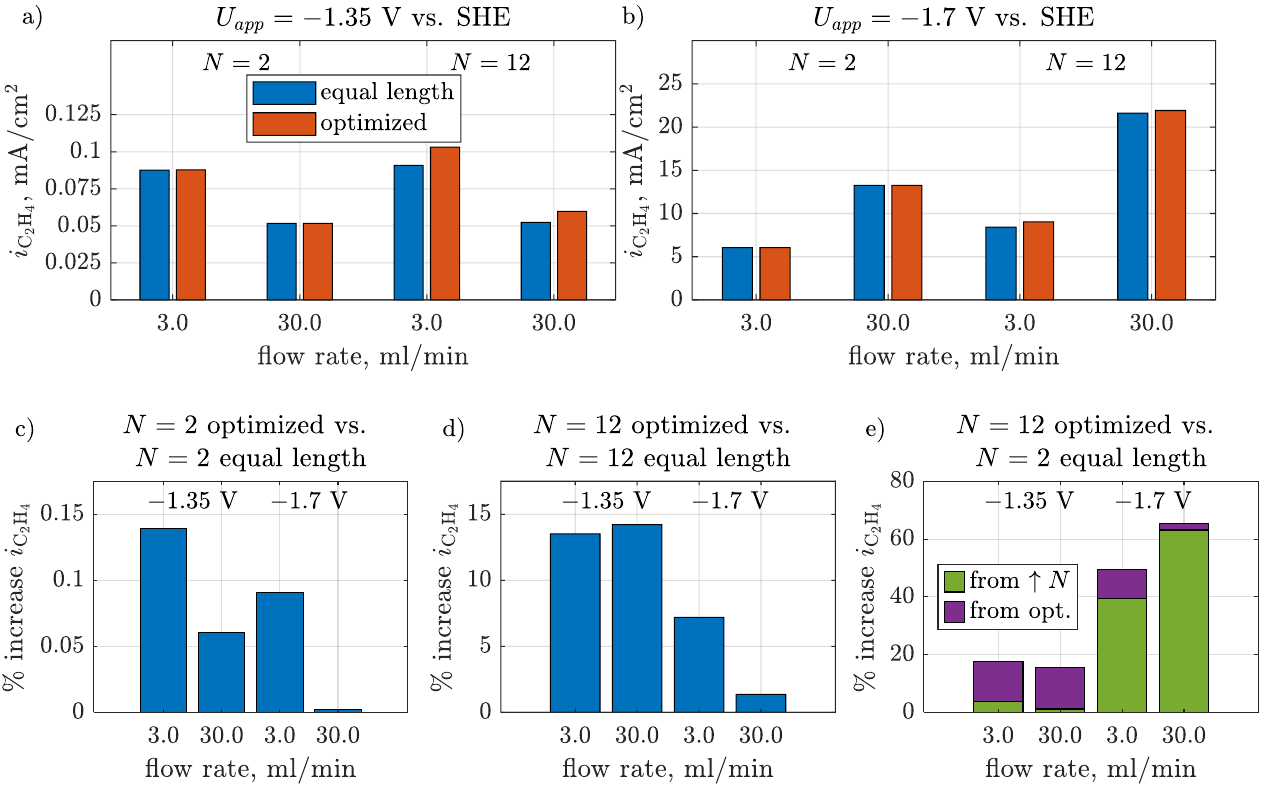}
    \caption{\doublespacing Plot of ethylene current density ($i_{\ce{C_2H_4}}$) values and percentage increases relative to several baseline comparison values. All optimized cases shown here use $\max (i_{\ce{C_2H_4}})$ as the objective function. a) $i_{\ce{C_2H_4}}$, in mA/cm\textsuperscript{2}, for the less-negative applied voltage condition: $U_{app} = -1.35$ V vs. SHE. b) $i_{\ce{C_2H_4}}$, in mA/cm\textsuperscript{2}, for the more-negative applied voltage condition: $U_{app} = -1.7$ V vs. SHE. c) Percentage increase in $i_{\ce{C_2H_4}}$ for the optimized compared to the equal length cases, for $N = 2$. d) Percentage increase in $i_{\ce{C_2H_4}}$ for the optimized compared to the equal length cases, for $N = 12$. e) Percentage increase in $i_{\ce{C_2H_4}}$ for the $N = 12$ optimized compared to the $N = 2$ equal length cases. The percentage increase is separated into two contributions: from the increase in $N$, and from the optimization.}  
\label{fig:current_density_plot}
\end{figure}

\Cref{fig:current_density_plot} shows the values of the ethylene current density $i_{\ce{C_2H_4}}$ for $N = 2$ and $N = 12$, across all combinations of applied voltage ($U_{app}$) and flow rate. From the values of $i_{\ce{C_2H_4}}$ in \cref{fig:current_density_plot}a and \cref{fig:current_density_plot}b, some key trends emerge:
\begin{enumerate}
    \item The values of $i_{\ce{C_2H_4}}$ are significantly larger (by around two orders of magnitude) for the $U_{app} = -1.7$ V vs. SHE cases compared to the $U_{app} = -1.35$ V vs. SHE cases.
    \item For a given $U_{app}$ and flow rate, increasing $N$ leads to increased $i_{\ce{C_2H_4}}$, and the relative amount of increase is larger at more negative $U_{app}$.
    \item Increased flow rates yield larger \ce{C_2H_4} current density values at the more negative $U_{app} = -1.7$ V vs. SHE, but yield smaller \ce{C_2H_4} current density values at the less negative $U_{app} = -1.35$ V vs. SHE.
\end{enumerate}

\Cref{fig:current_density_plot}c and \cref{fig:current_density_plot}d show the percentage increase in $i_{\ce{C_2H_4}}$, for $N = 2$ in c and for $N = 12$ in d, when comparing the optimized patterning cases compared to the equal length patterning cases. Minimal \% increase is achieved for $N = 2$ ($< 0.15$\% across all $U_{app}$ and flow rate values), while for $N = 12$ the \% increase values are significantly larger (approaching $15$\% at $U_{app} = -1.35$ V vs. SHE).

The observed benefits in electrochemical performance for the best-performing cases can be divided into two contributing factors: 1) increasing the number of sections $N$ and 2) optimizing their lengths. To emphasize and quantify these combined benefits of increased $N$ and optimization, \cref{fig:current_density_plot}e shows the \% increase in $i_{\ce{C_2H_4}}$ observed in the $N = 12$ optimized cases compared to the $N = 2$ equal length cases. All conditions yield significant percentage increases, with $\simeq 15 - 18$ \% at $U_{app} = -1.35$ V vs. SHE and $\simeq 50 - 65$ \% at $U_{app} = -1.7$ V vs. SHE. At $U_{app} = -1.35$ V vs. SHE, the optimization in patterning at the larger $N$ is the predominant contribution towards the overall percentage increase. At $U_{app} = -1.7$ V vs. SHE, the opposite is true, with the increase $N$ being by far the largest contribution.

\begin{figure}[!h]
    \centering
    \includegraphics[width=\textwidth]{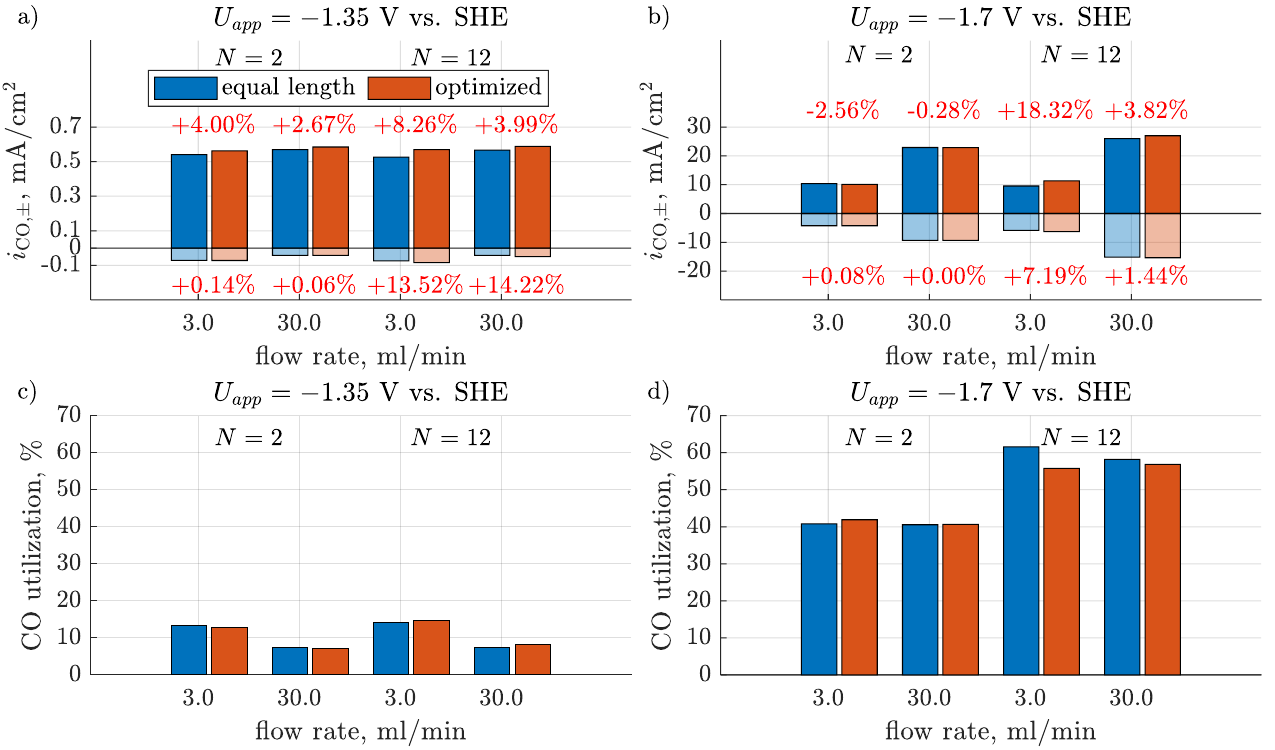}
    \caption{\doublespacing \ce{CO} production $i_{\ce{CO},+}$ and consumption $i_{\ce{CO},-}$, as calculated from \cref{eq:CO_production_consumption}, for the equal length case and optimized case of each condition, shown in a) for $U_{app} = -1.35$ V vs. SHE cases, and in b) for $U_{app} = -1.7$ V vs. SHE cases. For each case, the $i_{\ce{CO},-}$ value is multiplied by $-1$ and shown underneath the $i_{\ce{CO},+}$ bar, with the same color but lower opacity. Above each $i_{\ce{CO},+}$ pair of columns and below each $i_{\ce{CO},-}$ pair of columns, the percentage change of the optimized compared to the equal length value is shown. The \ce{CO} utilization per \cref{eq:CO_utilization} is shown in c) for $U_{app} = -1.35$ V vs. SHE cases, and in d) for $U_{app} = -1.7$ V vs. SHE cases.}
    \label{fig:CO_production_consumption}
\end{figure}

As the key intermediate species in tandem CO2R reaction systems, \ce{CO} warrants further analysis of its current density behavior. For this purpose, \cref{fig:CO_production_consumption}a and \cref{fig:CO_production_consumption}b show the net \ce{CO} current density $i_{\ce{CO},\text{net}}$, decomposed into the production $i_{\ce{CO},+}$ and consumption $i_{\ce{CO},-}$. Values are shown for the equal length case and the optimized case at all conditions for $N = 2$ and $12$. The expressions for $i_{\ce{CO},+}$ and $i_{\ce{CO},-}$ are given in \cref{eq:CO_production_consumption}.

It is observed that the magnitude of $i_{\ce{CO},-}$ in each optimized case is greater than or equal to the equal length case magnitude across all conditions and $N$ values. Because \ce{CO} is consumed in the production of \ce{C_2H_4} (see \cref{eq:surface_rxn_1_Cu}), increasing the \ce{CO} consumption is necessary towards increasing $i_{\ce{C_2H_4}}$ to satisfy the optimization objective function.

Towards increased understanding of this \ce{CO} current density data, a parameter can be constructed that measures the proportion of the produced \ce{CO} that is consumed by the \ce{Cu} surface reactions, i.e. 
\begin{equation}\label{eq:CO_utilization}
    \ce{CO}\text{ utilization} = \frac{i_{\ce{CO},-}}{i_{\ce{CO},+}}.
\end{equation}
\Cref{fig:CO_production_consumption}c and \cref{fig:CO_production_consumption}d show the \ce{CO} utilization values across the cases in this study. The less negative $U_{app} = -1.35$ V vs. SHE cases show utilization levels of less than $15$\% across all conditions, while the more negative $U_{app} = -1.7$ V vs. SHE cases show \ce{CO} utilization values that are all substantially higher (all between $40$\% and $62$\%). However, for all conditions and $N$ values as a whole, no significant increase (or decrease) in the \ce{CO} utilization is observed for the optimized case values compared to the equal length case values.

When observing how the \ce{CO} utilization depends on the flow rate, the behavior differs significantly at the two different $U_{app}$ values considered, as discussed below.

For all of the less negative $U_{app} = -1.35$ V vs. SHE cases, increasing the flow rate causes a significant decrease in the \ce{CO} utilization, with the $30.0$ ml/min case values decreasing to $53.4-56.7$\% of the corresponding $3.0$ ml/min case value. As seen in \cref{fig:CO_production_consumption}a, the decreased \ce{CO} utilization occurs because across all cases, for increased flow rate, $i_{\ce{CO},+}$ increases slightly (by up to $\simeq 8$\%) while $i_{\ce{CO},-}$ decreases significantly (by more than $57$\%). In short, despite a larger amount of produced \ce{CO} that enters the system into the bulk electrolyte at the higher flow rate, the system is only about half as effective as it was at the lower flow rate in utilizing the \ce{CO} using the \ce{Cu} surface reactions. The underlying mechanism for this effect becomes clear in \cref{sec:Analysis_concentration_fields} through observation of the combined interactions of the surface reaction strength and flow rate in the concentration fields.

Conversely, across all of the more negative $U_{app} = -1.7$ V vs. SHE cases, increasing the flow rate does not cause a significant decrease in \ce{CO} utilization. Values in the $30.0$ ml/min cases are within $94.5$\% of the corresponding $3.0$ ml/min case value. Additionally, as seen in panel b of \cref{fig:CO_production_consumption}, both the $i_{\ce{CO},+}$ and $i_{\ce{CO},-}$ more than double in value for $30.0$ ml/min compared to the corresponding value at $3.0$ ml/min. Thus, at the higher flow rate significantly more \ce{CO} is generated, and even with this increased amount of \ce{CO} the overall modeled reactor remains nearly as effective as at the lower flow rate in consuming the available \ce{CO} via the \ce{Cu} surface reactions. Just as was mentioned for the less negative $U_{app}$, this effect is explained by the impacts of surface reaction strength and flow rate in the concentration fields, as shown and discussed in \cref{sec:Analysis_concentration_fields}.

Tying these back to the reactor performance, increasing (or decreasing) the \ce{CO} consumption directly contributes to an increase (or decrease) in $i_{\ce{C_2H_4}}$, due to \ce{CO} being the reactant species for the \ce{C_2H_4} electrochemical reaction. Thus, when considering the correlations of $i_{\ce{C_2H_4}}$ with flow rate as seen in \cref{fig:current_density_plot}, the inverse correlation at $U_{app} = -1.35$ V vs. SHE and the direct correlation at $U_{app} = -1.7$ V vs. SHE are both consistent with the \ce{CO} consumption trends described here.

\subsection{Analysis of concentration fields}\label{sec:Analysis_concentration_fields}

The electrochemical performance results and discussion in \cref{sec:Current_density_performance} focus on the electrode-averaged current density (i.e., the system-level performance). In this section, the concentration field results are presented, in order to gain insight into how the local environment features determine the overall flow reactor performance.

The first results presented in this section are the $N = 2$ equal length cases, in \cref{sec:N_2_equal_length_cases}. The goal is to isolate the key processes fundamental to the tandem reaction setup, which thus can be studied even in this simplest configuration of one \ce{Ag} and and one \ce{Cu} section. Following this, the $N = 12$ results are presented in \cref{sec:N_12_cases}, for both the optimized and the equal length patterning cases. With these more complicated cases, one introduced effect is that with increased patterning (with larger $N$), the \ce{Ag} sections located further downstream from the leading edge can utilize any non-reacted \ce{CO_2}, thus producing more \ce{CO} to be used by \ce{Cu} sections also located downstream. Another introduced effect is that the optimized cases yield non-uniform \ce{Ag} and \ce{Cu} section lengths, causing more complicated species boundary layers growths and interactions. Understanding these effects that arise from increased patterning and optimized designs is important, as these cases yield noticeable increases in the overall current density performance (as seen in \cref{sec:Current_density_performance}).

\subsubsection{Unoptimized equal length cases, \texorpdfstring{$N = 2$}{N = 2}}\label{sec:N_2_equal_length_cases}

\begin{figure}[!h]
    \centering
    \includegraphics[width=\textwidth]{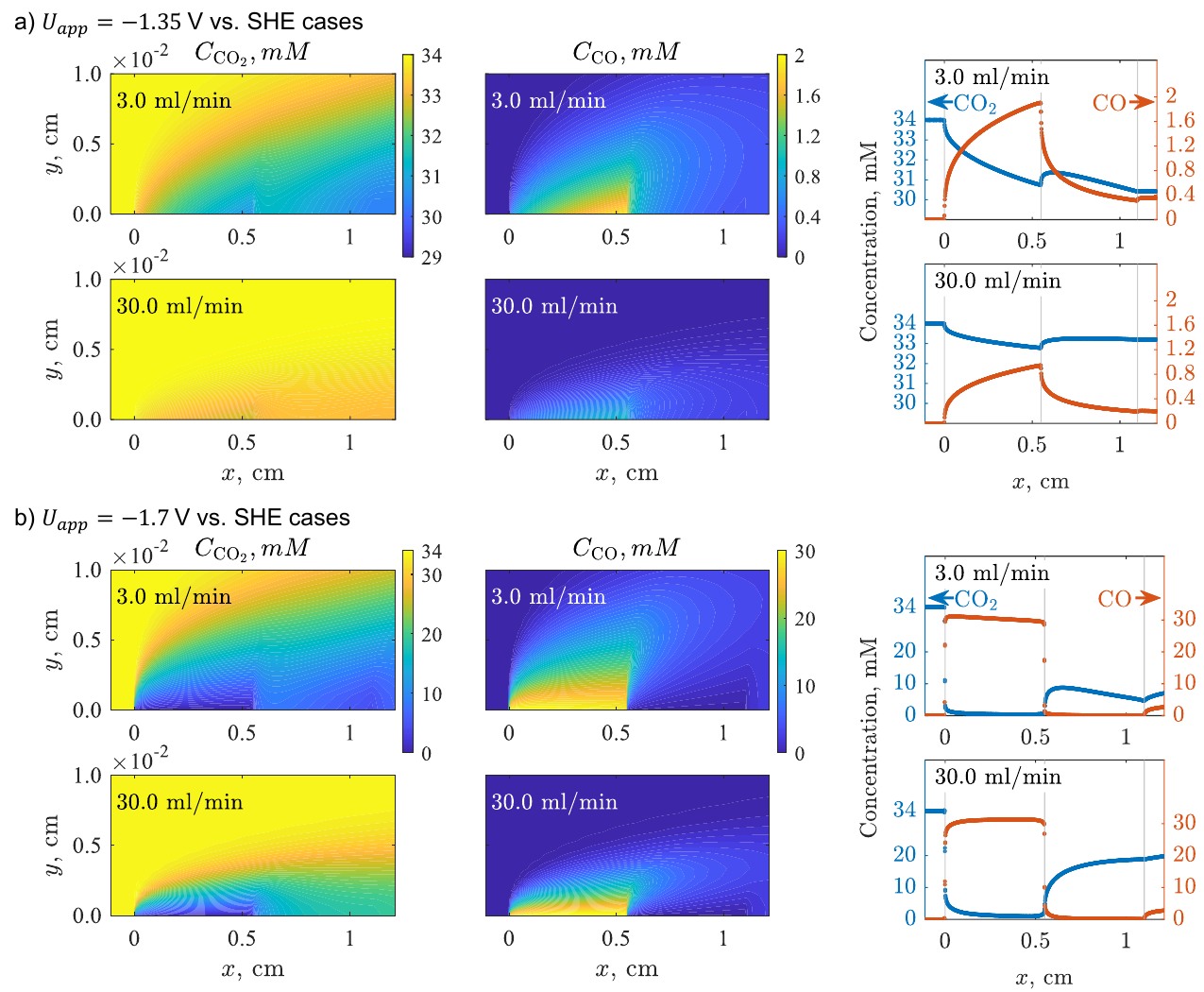}
    \caption{\doublespacing Concentration field contour plots and surface concentration line plots for \ce{CO_2} and \ce{CO}, for the equal length patterning with $N = 2$. Plots are shown for a) the less negative $U_{app} = -1.35$ V vs. SHE applied voltage, and for b) the more negative $U_{app} = -1.7$ V vs. SHE applied voltage. For the surface concentrations plots, vertical gray lines denote the locations of the section boundaries.}
    \label{fig:concfield_highlowUapp_N2}
\end{figure}

\Cref{fig:concfield_highlowUapp_N2} shows the contours and surface concentration profiles of \ce{CO_2} and \ce{CO} for the $N = 2$ equal length cases. As seen in both the contours and the surface concentration profiles, the regions of low \ce{CO_2} and high \ce{CO} concentrations clearly delineate the $x$-boundaries of the \ce{Ag} section.

In the surface concentration profiles, the sharpness of the concentration gradients ($dc_k/dx$) at the section boundaries provides a sense of the relative surface reaction strength, with the $U_{app} = -1.7$ V vs. SHE gradients all being much sharper (corresponding to the stronger reaction rates for both \ce{Ag} and \ce{Cu}). It is also observed that large proportions of \ce{Cu} sections near the downstream trailing edge exhibit low \ce{CO} surface concentrations; these ``dead zones'' yield only minimal production of \ce{Cu} products. It is shown and discussed in \cref{sec:N_12_cases} that reducing the existence and impact of these dead zones is a key hypothesis of how the optimized designs maximize the \ce{CO} consumption to maximize the ethylene current density objective.

The $U_{app} = -1.35$ V vs. SHE contours and profiles exhibit a strong dependency on flow rate, with the higher flow rate showing significantly larger \ce{CO_2} concentrations and smaller \ce{CO} concentrations. On the other hand, the $U_{app} = -1.7$ V vs. SHE plots show only minor differences between the low and high flow rate plots. These differences arise from how \ce{CO_2} and \ce{CO} each enter the system, combined with the relative strengths of the flow rate compared to the surface reactions; these behaviors are discussed here below.

\underline{\ce{CO_2}}: Because \ce{CO_2} is fed into the system at the inlet at $34$ mM for each case, its supply is always guaranteed to reach the surface of the first \ce{Ag} section. Regardless of $U_{app}$, increasing the flow rate more effectively replenishes high concentration \ce{CO_2} fluid onto the \ce{Ag} section surface, which increases the \ce{CO} production (as reflected in \cref{fig:CO_production_consumption}).

\underline{\ce{CO}}: On the other hand, \ce{CO} is not supplied into the system at the inlet, as it must be created through the \ce{Ag} surface reaction. This ``non-guaranteed'' supply means that its behavior in response to flow rate depends on the relative surface reaction strength.

At $U_{app} = -1.35$ V vs. SHE, the weaker surface reaction strength is overcome by the flow strength; thus, the flow sweeps the \ce{CO} downstream such that high concentrations are unable to accumulate above the \ce{Ag} section surface. Thus, the already limited supply of \ce{CO} that reaches \ce{Cu} surface is made even more scarce at high flow rates. This behavior mechanistically explains why the increased flow rate causes a decrease in \ce{CO} consumption as seen in \cref{fig:CO_production_consumption} and thus leads to the decrease in $i_{\ce{C_2H_4}}$ as seen in \cref{fig:current_density_plot}.

At $U_{app} = -1.7$ V vs. SHE, the stronger surface reaction strength overcomes the flow strength, so that high \ce{CO} concentrations are able to accumulate on the \ce{Ag} for both slow and fast flow rates. The high \ce{CO} concentrations that are generated on the \ce{Ag} surface ``guarantees'' its supply to reach the \ce{Cu} section, much like the \ce{CO_2} for \ce{Ag}. Increasing the flow rate thus increases the replenishment of high \ce{CO} to reach the \ce{Cu} surface, contributing to the increased \ce{CO} consumption as seen in \cref{fig:CO_production_consumption} and thus leading to the increased $i_{\ce{C_2H_4}}$ as seen in \cref{fig:current_density_plot}.

As a note, as seen in \cref{fig:concfield_highlowUapp_N2}, the surface concentration of $\ce{CO}$ can exceed the $0.95$ mM solubility limit. This gas supersaturation near the electrode surface is a well-documented effect, especially for smooth electrode surfaces that are capable of suppressing bubble nucleation \citep{lum2018sequential}.

\subsubsection{Equal length and optimized cases, \texorpdfstring{$N = 12$}{N = 12}}\label{sec:N_12_cases}

\begin{figure}[!h]
    \centering
    \includegraphics[width=\textwidth]{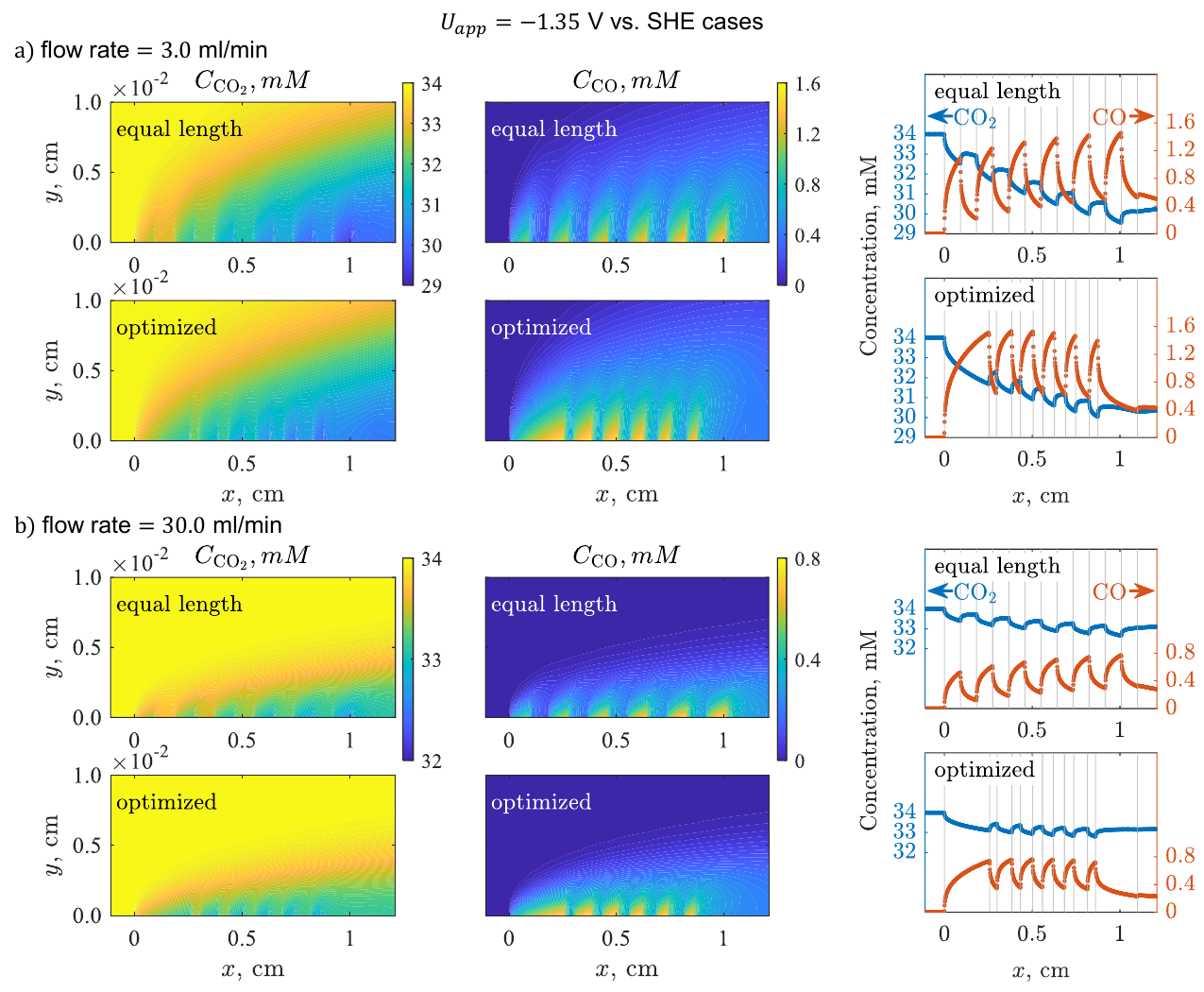}
    \caption{\doublespacing Concentration field contour plots and surface concentration line plots for \ce{CO_2} and \ce{CO}, for both the equal length and optimized patternings for the less negative $U_{app} = -1.35$ V vs. SHE applied voltage. Plots are shown for a) the lower $3.0$ ml/min flow rate cases, and for b) the higher $30.0$ ml/min flow rate cases. For the surface concentrations plots, vertical gray lines denote the locations of the section boundaries.}
    \label{fig:concfield_lowUapp}
\end{figure}

Armed with intuition from the $N=2$ cases, we are now prepared to analyze the more complex $N=12$ cases. \Cref{fig:concfield_lowUapp} shows the concentration field contours and surface concentration profiles for the $N = 12$ cases at $U_{app} = -1.35$ V vs. SHE. As is also observed in the $N = 2$ equal length cases, the \ce{CO_2} concentration stays high along the electrode surface (due to the limited consumption by the weak \ce{Ag} surface reaction), especially at the higher flow rate. An effect that is only seen with larger $N$ values, in the equal length cases at each flow rate, is that the \ce{CO} surface concentration steadily increases moving downstream, as the \ce{CO} that was not consumed from the preceding upstream \ce{Cu} section combines with the \ce{CO} production of each subsequent \ce{Ag} section to accumulate on that \ce{Ag} section.

Now shifting focus towards the optimized designs, the design patternings (as also shown in \cref{fig:opt_vs_N_figure}) feature long lengths for the first \ce{Ag} and the last \ce{Cu} sections, with alternating short sections in between. For the optimized case for each flow rate in \cref{fig:concfield_lowUapp}, the first long \ce{Ag} section yields a noticeably higher peak surface \ce{CO} concentration than in the corresponding equal length case. The shorter intermediate \ce{Ag} and \ce{Cu} sections then work together to preserve the surface \ce{CO} concentration at a higher minimum value and higher average value overall than in the corresponding equal length case. Compared to the values for the corresponding equal length cases, the increased \ce{CO} concentration for the optimized cases reflects the higher overall \ce{CO} production, which also yields increased reactant supply for the \ce{Cu} surface reactions and ultimately leads to increased \ce{CO} consumption towards satisfying the objective of maximizing \ce{C_2H_4} current density. The increased \ce{CO} production and consumption for the optimized cases at these $N = 12, U_{app} = -1.35$ V vs. SHE conditions are consistent with the data shown in \cref{fig:CO_production_consumption}.  Finally, the final long \ce{Cu} section consumes as much of the remaining \ce{CO} as possible as the fluid moves towards the domain exit.

Note that the \ce{CO} surface concentration in the equal length cases steadily increases moving downstream, eventually reaching similarly high maximum values to those attained in the corresponding optimized cases. However, the overall reactor performance is measured by the integral-average \ce{C_2H_4} current density, which is why maximizing this parameter is the objective function in the problem formulation (\cref{eq:opt_problem}). Therefore, attaining a high \ce{CO} surface concentration as close as possible to the upstream catalyst edge and then maintaining it at a high value provides larger overall reactant availability to then allow for increased \ce{C_2H_4} production, thus explaining the improved performance of the optimized cases.

\begin{figure}[!h]
    \centering
    \includegraphics[width=\textwidth]{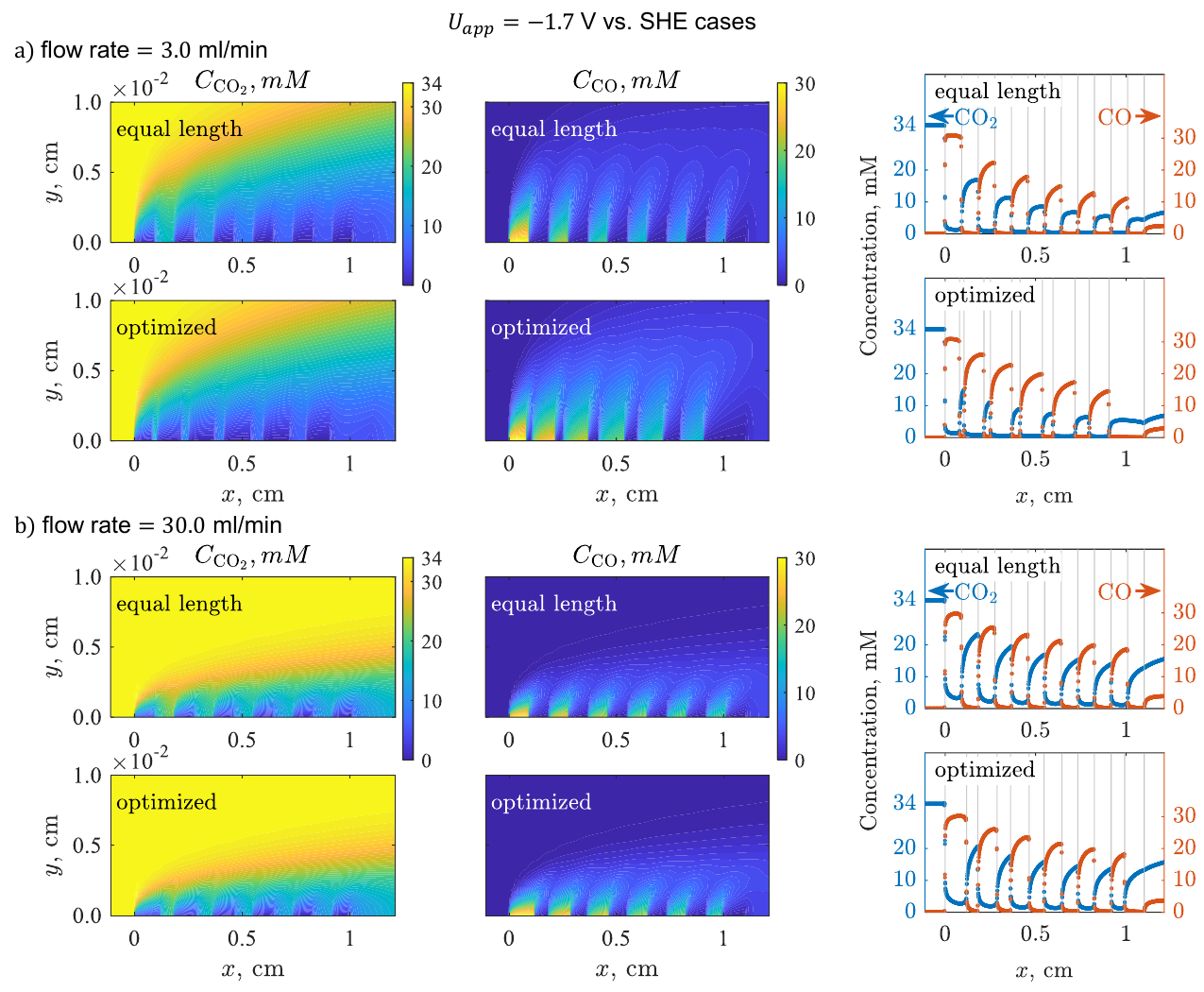}
    \caption{\doublespacing Concentration field contour plots and surface concentration line plots for \ce{CO_2} and \ce{CO}, for both the equal length and optimized patternings for the more negative $U_{app} = -1.7$ V vs. SHE applied voltage. Plots are shown for a) the lower $3.0$ ml/min flow rate cases, and for b) the higher $30.0$ ml/min flow rate cases. For the surface concentrations plots, vertical gray lines denote the locations of the section boundaries.}
    \label{fig:concfield_highUapp}
\end{figure}

Now moving to the more negative applied voltage, \cref{fig:concfield_highUapp} shows the concentration field contours and surface concentration profiles for the $N = 12$ cases at $U_{app} = -1.7$ V vs. SHE. In these cases, as is also observed in the $N = 2$ equal length cases, the strong strength of the surface reactions causes the \ce{Ag} sections to consume significant \ce{CO_2} (concentration $\rightarrow 0$) and produce significant \ce{CO}, and also causes the \ce{Cu} sections to consume significant \ce{CO} (concentration $\rightarrow 0$). As expected, at the faster flow rate, the reactant concentrations are higher due to the increased mass transport, which causes high reactant concentration fluid to be transported to the downstream catalyst sections' surfaces.

For the equal length cases at both flow rates, the surface concentration of \ce{CO} on the \ce{Cu} sections consistently reaches a lower value than that of \ce{CO_2} on the \ce{Ag} sections. While visible from the surface concentration plots, this comparison is made clearer using plots of the averaged surface concentrations in section S8 of the Supplementary Information.
Thus, it is hypothesized that in the equal length patterns, the ``dead zones'' of low \ce{CO} reactant on \ce{Cu} sections are more prevalent and have a larger impact in reducing the system performance compared to the ``dead zones'' of low \ce{CO_2} reactant on \ce{Ag} sections. 

We now turn our focus to the patterning design and the resulting concentration behavior of the optimized cases. As is pointed out in \cref{sec:Optimal_patterns}, the optimized patterns for these $U_{app} = -1.7$ V vs. SHE cases feature longer \ce{Ag} sections at the expense of shortened \ce{Cu} sections, especially for the low flow rate case. By shortening the \ce{Cu} sections, the optimized designs minimize the more detrimental negative impact of the \ce{CO} ``dead zones'', opting to allocate the unused length to create longer \ce{Ag} sections to yield increased \ce{CO} production. This leads to increased reactant \ce{CO} concentration on the \ce{Cu} sections, which thus leads to increased \ce{CO} consumption. In summary, our interpretation of the optimization algorithm's behavior at the stronger surface reaction conditions is that the lost consumption from removing the ``dead zone'' \ce{Cu} regions is outweighed by increasing the \ce{Ag} section lengths to produce more \ce{CO}. Overall, this leads to increased \ce{CO} consumption and increased $i_{\ce{C_2H_4}}$, thus satisfying the objective function.

\section{Conclusions}\label{sec:Conclusions}

Towards exploring design principles for higher performance CO2R electrolyzers, this work presents a computational simulation framework that combines continuum transport modeling, tandem catalysis, and design optimization in a flow reactor case study. In the tandem catalyst patterning, \ce{Ag} sections provides the \ce{CO_2 -> CO} reaction capability, and \ce{Cu} sections  provides the \ce{CO ->} high-value products reaction capability.

A set of cases is run to study the effects of changing the following input parameters: the surface applied voltage $U_{app}$, the electrolyte flow rate, and the number of sections $N$. Cases are run for equal length patterns and for optimized patterns (which maximize the current density towards ethylene, $i_{\ce{C_2H_4}}$, as the objective function). For the resultant optimized section patterns, less negative $U_{app}$ (weaker surface reaction) conditions favor longer first and last sections with shorter intermediate sections. Optimized patterns for more negative $U_{app}$ (stronger surface reaction) conditions are largely more uniform in section length but show a flow rate dependency.

The reactor performance, assessed using $i_{\ce{C_2H_4}}$, demonstrates a \% increase for the optimized cases that can be decomposed into two contributing components: 1) from increased $N$, and 2) from optimization.
The contribution from optimization dominates the \% increase at less negative $U_{app}$, while the contribution from increased $N$ dominates at more negative $U_{app}$. The highest performance increases in this study occur at the more negative $U_{app} = -1.7$ V vs. SHE value: using $N = 12$ sections, \% increases in $i_{\ce{C_2H_4}}$ of up to $65$\% are achieved compared to the corresponding unoptimized equal-length $N = 2$ cases. 
Analysis of the concentration fields provides mechanistic insight to explain the final optimized patterns. The key goal of the optimized patterns is to minimize regions of low \ce{CO} concentration on the \ce{Cu} section surfaces, thus maximizing \ce{C_2H_4} production. Because \ce{CO} is the reaction system's key intermediate species, trends in \ce{CO} production and consumption closely align with the concentration field observations.

This work demonstrates that the combination of informed continuum modeling with high performance computing (HPC) tools to drive design optimization provides a powerful method for studying electrochemical processes. Future directions of work include investigating other reaction systems, investigating other design optimization objective functions (such as selectivity or energy efficiency metrics), or modifying the setup to allow for optimization of operating conditions including $U_{app}$ and flow rate. Another potential future direction is to extend the developed optimization capability to three-dimensional (3D) settings, such as gas diffusion electrodes (GDEs), modifications of the flow field, and other complex cell configurations. Forward analysis of 3D setups in previous studies, such as modified \ce{Ag}/\ce{Cu} catalyst layer segment lengths and loadings in GDEs \citep{zhang2022highly}, have yielded complementary findings to the insights from the current study; thus, applying design optimization to such configurations is a logical next step for future studies. For 3D settings, incorporating shape/topology optimization capabilities for electrode morphology modification also comes into play. In experimental settings, the capabilities presented in this work help narrow the space of possible designs to focus on parameter choices and setups with high performance promise.

\section*{Supporting Information:}
Supplementary Information document with parameter values, modeling details, and supplementary results (PDF).

A branch of EchemFEM with scripts (and documentation) to run the optimization cases presented in this work has been made available. The URL for this link is: \url{https://github.com/llnl/echemfem/tree/spatial_patterning_opt}.

\section{Acknowledgments}
This work was performed under the auspices of the U.S. Department of Energy by Lawrence Livermore National Laboratory (LLNL) under Contract No. DE-AC52-07NA27344, and was partially supported by a Cooperative Research and Development Agreement (CRADA) between LLNL and TotalEnergies American Services, Inc. (affiliate of TotalEnergies SE) under Agreement No. TC02307. This work was supported by the Korea Institute of Energy Technology Evaluation and Planning (KETEP) and the Ministry of Trade, Industry \& Energy (MOTIE) of the Republic of Korea (RS-2024-00488176) through funding to J.-W.J. LLNL release number: LLNL-JRNL-2012882.

\section*{ORCID iDs}
\begin{enumerate}
    \item Jack Guo: 0000-0003-4090-9289
    \item Thomas Roy: 0000-0002-4286-4507
    \item Nitish Govindarajan: 0000-0003-3227-5183
    \item Joel B. Varley: 0000-0002-5384-5248
    \item Jonathan Raisin: 0009-0009-6455-599X
    \item Jinyoung Lee: 0000-0002-6905-896X
    \item Jiwook Jang: 0000-0003-1251-1011
    \item Dong Un Lee: 0000-0001-7591-5350
    \item Thomas F. Jaramillo: 0000-0001-9900-0622
    \item Tiras Y. Lin: 0000-0002-3377-9933
\end{enumerate}


\begingroup
\bibliographystyle{unsrtnat}
\bibliography{references}
\endgroup



\end{document}



\begin{frontmatter}

\title{Supplementary information\\\normalsize Optimized tandem catalyst patterning for CO$_2$ reduction flow reactors}



\author[address1]{Jack Guo\corref{mycorrespondingauthor}}
\cortext[mycorrespondingauthor]{Corresponding author}
\ead{guo9@llnl.gov}

\author[address1]{Thomas Roy}

\author[address2]{Nitish Govindarajan}

\author[address2]{Joel B. Varley}

\author[address3,address5]{Jonathan Raisin}

\author[address3,address4,address6]{Jinyoung Lee}

\author[address6,address7]{Ji-Wook Jang}

\author[address3,address4]{Dong Un Lee}

\author[address3,address4]{Thomas F. Jaramillo}

\author[address1]{Tiras Y. Lin\corref{mycorrespondingauthor}}
\ead{lin46@llnl.gov}

\address[address1]{Computational Engineering Division, Lawrence Livermore National Laboratory, Livermore, CA 94550, USA}

\address[address2]{Materials Science Division, Lawrence Livermore National Laboratory, Livermore, CA 94550, USA}

\address[address3]{SUNCAT Center for Interface Science and Catalysis, Department of Chemical Engineering, Stanford University, Stanford, CA 94305, USA}

\address[address4]{SUNCAT Center for Interface Science and Catalysis, SLAC National Accelerator Laboratory, Menlo Park, CA 94025, USA}

\address[address5]{TotalEnergies Research \& Technology USA LLC, Houston, TX 77002, USA}

\address[address6]{School of Energy and Chemical Engineering, Ulsan National Institute of Science and Technology (UNIST), Ulsan, Ulsan, 44919, Republic of Korea}

\address[address7]{Graduate School of Carbon Neutrality, Ulsan National Institute of Science and Technology (UNIST), Ulsan, Ulsan, 44919, Republic of Korea}

\end{frontmatter}

\tableofcontents

\clearpage

\section{Tabulated values}\label{sec:Tabulated_values}
\bgroup
\def\arraystretch{1.1}

\begin{table}[!h]
\centering
\caption{\doublespacing Table of diffusivity values for each chemical species.}
\begin{tabular}{|l|l|l|}
\hline
Diffusivity     & Value (m$^2$ s$^{-1}$)   & Ref.                                    \\
\hline
$D_{ \ce{CO_2} }$      & $1.91 \times 10^{-9}$  & \cite{weng2018modeling}     \\

$D_{ \ce{K^+} }$       & $1.957 \times 10^{-9}$  & \cite{weng2018modeling}                                             \\

$D_{ \ce{OH^-} }$      & $5.29 \times 10^{-9}$  & \cite{weng2018modeling}                                             \\ 

$D_{ \ce{H^+} }$       & $9.311 \times 10^{-9}$ &   \cite{weng2018modeling}                                           \\

$D_{ \ce{CO_3^{2-}} }$ & $0.92 \times 10^{-9}$  & \cite{weng2018modeling}                                             \\ 

$D_{ \ce{HCO_3^-} }$   & $1.185 \times 10^{-9}$ & \cite{weng2018modeling}                                             \\

$D_{ \ce{CO} }$        & $2.03 \times 10^{-9}$  & \cite{cussler2009diffusion} \\

$D_{ \ce{H_2} }$       & $4.50 \times 10^{-9}$   & \cite{cussler2009diffusion}  \\
$D_{ \ce{C_2H_4} }$       & $1.87 \times 10^{-9}$   &  \cite{cussler2009diffusion} \\
$D_{ \ce{C_2H_6O} }$       & $0.84 \times 10^{-9}$   &  \cite{cussler2009diffusion}  \\
$D_{ \ce{CH_4} }$       & $1.49 \times 10^{-9}$   &  \cite{cussler2009diffusion} \\
\hline 
\end{tabular}

\label{table:Diffusivity}
\end{table}

\begin{table}[!h]
\centering
\caption{\doublespacing Table of reaction rate constant values for the reactions in the bulk bicarbonate electrolyte (as listed in eq. (1) in the main text). 
For bulk reaction index $j$, certain values are provided in terms of the equilibrium constant $K_j$ and the dissociation constant of water $K_w$.}
\begin{tabular}{|l|l|l| |l|l|l|}
\hline
Rxn. rate & Value \& Units & Ref. & Rxn. rate & Value \& Units & Ref.     \\
\hline
$K_{1}$ & $10^{-6.37}$ mol L$^{-1}$ & \citep{weng2018modeling}   & $k_{2f}$  & $3.71 \times 10^{-2}$ s$^{-1}$  &  \citep{weng2018modeling}   \\
$K_{2}$ & $10^{-10.32}$ mol L$^{-1}$ & \citep{weng2018modeling}  & $k_{2r}$  & $k_{2f} / K_1$    &   \citep{weng2018modeling}  \\
$K_{w}$ & $10^{-14}$ mol$^2$ L$^{-2}$ & \citep{weng2018modeling}  & $k_{3f}$ & $k_{3r} / K_4$  & \citep{weng2018modeling} \\
$K_{3}$ & $K_1 / K_w$ & \citep{weng2018modeling} & $k_{3r}$  & $6.0 \times 10^9$ L mol$^{-1}$ s$^{-1}$  &  \citep{weng2018modeling}   \\
$K_{4}$ & $K_2 / K_w$ & \citep{weng2018modeling} & $k_{4f}$ & $k_{4r} / K_2$ &   \citep{weng2018modeling} \\
$k_{1f}$ & $8.42 \times 10^3$ L mol$^{-1}$ s$^{-1}$ & \citep{astarita1983gas}  & $k_{4r}$ & $59.44$ s$^{-1}$ & \citep{weng2018modeling}  \\
$k_{1r}$ & $k_{1f} / K_3$ & \citep{weng2018modeling} & $k_{5f}$ & $2.3 \times 10^{10}$ L mol$^{-1}$ s$^{-1}$  & \citep{schulz2006determination} \\
& & & $k_{5r}$  & $K_w k_{4f}$   & \citep{weng2018modeling}\\
\hline
\end{tabular}

\label{table:rxn_rate_const}
\end{table}

\begin{table}[!h]
\centering
\caption{\doublespacing Table of Sechenov coefficients, used in the surface reaction modeling for \ce{Ag}.}
\begin{tabular}{|l|l|l|}
\hline
Sechenov constant & Value (m$^3$ mol$^{-1}$)                                   & Ref.                      \\
\hline
$h_{g, \ce{CO_2} }$               & $-1.7159 \times 10^{-5}$      & \cite{zeng2020kinetic}         \\
$h_{s, \ce{H^+} }$               & $0.0$      & \cite{zeng2020kinetic}         \\
$h_{s, \ce{OH^-} }$               & $8.39 \times 10^{-5}$      & \cite{zeng2020kinetic}         \\
$h_{s, \ce{K^+} }$               & $9.2 \times 10^{-5}$      & \cite{zeng2020kinetic}         \\
$h_{s, \ce{CO_3^{2-}} }$               & $14.23 \times 10^{-5}$      & \cite{zeng2020kinetic}         \\
$h_{s, \ce{HCO_3^{-}} }$               & $9.67 \times 10^{-5}$      & \cite{zeng2020kinetic}         \\
\hline
\end{tabular}

\label{table:Sechenov_consts_for_activity}
\end{table}

\clearpage

\section{Fluid flow modeling details}\label{sec:Fluid_flow_modeling_details}

In flow reactor setups, the relative strengths of the dissolved species transport (i.e., mass transfer) effects versus fluid viscous effects at the catalyst surface can be quantified using the Schmidt number, which for each species $k$ is written as $Sc_k = \nu / D_k$, with $\nu$ being the fluid solvent viscosity. Values of $Sc_k$ are typically large ($>> 1$) in flow reactors \citep{lin2021analysis}, meaning that the concentration boundary layer of each species is much thinner than the momentum boundary layer. Thus, when modeling the near-wall flow reactor behavior, the fluid velocity can be approximated as a shear flow with constant shear rate $\dot{\gamma}$ as a simplified representation. The flow velocity is an input to the bulk chemistry through the species flux in eq. (4) of the main text; we employ a one-way coupled assumption to treat the flow velocity as being not influenced by the chemistry.

To select values of $Pe$, an example experimental 3D rectangular flow channel with electrode length $L_x = 2.2$ cm, width $w = 3.5$ mm, and height $h = 2$ mm is considered. The 2D shear rate of the computational domain can be analytically converted to and from the experimental flow rate of a 3D rectangular flow channel. Following \citet{stone2007introduction}, for a pressure-driven flow in a channel with a rectangular cross-section with flow rate $Q$, pressure drop $\Delta p$, and dynamic viscosity $\mu$, the pressure gradient is given as
\begin{equation}
    \frac{\Delta p}{L_x} = \frac{12 \mu Q}{w h^3} \frac{1}{1 - 6 \left( \frac{h}{w} \right) \sum_{n = 0}^\infty \lambda_n^{-5} \tanh \left( \frac{\lambda_n w}{h} \right)},
\end{equation}
where $\lambda_n = \left(2 n + 1 \right) \pi / 2$. The shear rate $\dot{\gamma}$, along the midline of the plate parallel to the flow direction, is given as
\begin{equation}
    \dot{\gamma} = \frac{\partial u}{\partial y} \bigg|_{\text{midline}} = \frac{\Delta p}{L_x} \frac{1}{2 \mu} \left( h + \frac{2}{h} \sum_{n = 0}^{\infty} a_n \lambda_n \sin(-\lambda_n) \right),
\end{equation}
where $a_n = h^2 (-1)^n / \left[ (\lambda_n)^3 \cosh(\lambda_n w / h) \right]$ are the Fourier coefficients solved using the side-wall no-slip boundaries. Thus, for an experimental (3D) flow rate $Q$, the shear rate $\dot{\gamma}$ can be computed and then plugged into eq. (22) of the main text to obtain the P\'eclet number $Pe$ needed for the computational simulation (2D) setup. This value of $\dot{\gamma}$ also determines the fluid velocity, as expressed in eq. (6) of the main text.

For this study, the conversion factor ends up as $Q = 1.0$ ml/min $\leftrightarrow$ $Pe = 2.41 \times 10^6$.

\section{\texorpdfstring{\ce{Ag}}{Ag} surface chemistry formulation}\label{sec:Ag_surface_chemistry_formulation}

The Tafel expressions\citep{corpus2023coupling} describe the \ce{Ag} surface reactions (given by eq. (10) and eq. (11) in the main text).
The Tafel expressions are re-written here:
\begin{equation}\label{eq:i_CO_Tafel}
\begin{split}
    i_{ \ce{CO}, \text{local} } = \left( \frac{a_{ \ce{ CO_2} }}{a_{ \ce{CO_2} }^{bulk}} \right)^{-\gamma_{ \ce{CO_2},\ce{CO} }} \left( \frac{a_{ \ce{OH^-} }}{a_{ \ce{OH^-} }^{bulk}} \right)^{\gamma_{ \ce{OH^-},\ce{CO} }} i_{0,\ce{CO}} \left( - \exp\left[ - \frac{\alpha_{c,\ce{CO}} F}{RT} \eta_{s,\ce{CO}} \right] \right),
\end{split}
\end{equation}
\begin{equation}\label{eq:i_H2_Tafel}
\begin{split}
    i_{\ce{H_2}, \text{local}} = \left( \frac{a_{ \ce{OH^-} }}{a_{ \ce{OH^-} }^{bulk}} \right)^{\gamma_{ \ce{OH^-},\ce{H_2} }} i_{0, \ce{H_2} } \left( - \exp\left[ - \frac{\alpha_{c, \ce{H_2} } F}{RT} \eta_{s, \ce{H_2} } \right] \right).
\end{split}
\end{equation}

The following exponents of the activity terms are defined:
\begin{subequations}
    \begin{align}
        \gamma_{ \ce{CO_2}, \ce{CO} } = -\frac{2 - \alpha_{c, \ce{CO} }}{2},\\
        \gamma_{ \ce{OH^-}, \ce{CO} } = \alpha_{c, \ce{CO} },\\
        \gamma_{ \ce{OH^-},\ce{H_2} } = \alpha_{c, \ce{H_2} }.
    \end{align}
\end{subequations}
and the following overpotential expressions are used:
\begin{subequations}
    \begin{align}
        \eta_{s, \ce{CO} } = \phi_s - \phi_l - \left( U_{ \ce{CO} }^0 - \frac{2.303 RT}{F} \text{pH} + \frac{RT}{2F} \ln \left( a_{ \ce{CO_2} } \right) \right),\\
        \eta_{s, \ce{H_2} } = \phi_s - \phi_l - \left( U_{ \ce{H_2} }^0 - \frac{2.303 RT}{F} \text{pH} \right).
    \end{align}
\end{subequations}
$\phi_s$ is the potential of the solid surface at $y = 0$, and is equal to the applied voltage $U_{app}$. In these expressions, the following values (as listed in \citet{corpus2023coupling}) of the transfer coefficient $\alpha_{c,\ce{k}}$, exchange current density $i_{0,\ce{k}}$, and reference voltage $U_{\ce{k}}^0$ are used:
\begin{equation}
    \begin{split}
        \alpha_{c, \ce{H_2} } = 0.312, \\
        \log_{10} \left( i_{0, \ce{H_2} } \left[ \text{mA cm}^{-2} \right] \right) = -6.77, \\
        \alpha_{c, \ce{CO} } = 0.544, \\
        \log_{10} \left( i_{0, \ce{CO} } \left[ \text{mA cm}^{-2} \right] \right) = -6.72,  \\
        U_{ \ce{CO} }^0 = -0.11 \text{ V versus SHE}, \\
        U_{ \ce{H_2} }^0 = 0.0 \text{ V versus SHE}. \\
    \end{split}
\end{equation}
The pH is defined as
\begin{equation}
    \text{pH} = -\log_{10} \left( a_{\ce{H^+} } \right),
\end{equation}
where $a_k$ refers to the activity of species $k$ in the model and is calculated as
\begin{equation}
    a_k = \frac{f_k c_k}{c_{ref}},
\end{equation}
where $c_{ref}$ is a reference concentration equal to $1 M$. $f_k$ is the activity coefficient of species $k$. For each ionic species, $f_k$ is calculated by the Davies activity coefficient, written as
\begin{equation}
    f_k = 10^{-0.51 z_k^2 \left( \frac{\sqrt{I}}{1 + \sqrt{I}} - 0.3 I\right)},
\end{equation}
where $I$ is the (molar) ionic strength of the electrolyte, which is defined as
\begin{equation}
    I = \frac{1}{2} \sum_k^N c_k z_k^2.
\end{equation}
Note that when computing $I$ to obtain $f_k$, the concentrations $c_k$ should be given in units of mol/L = M.

For \ce{CO_2}, the activity coefficient is given using Sechenov coefficients (whose values are provided in \cref{table:Sechenov_consts_for_activity}), as 
\begin{equation}
    f_{k = \ce{CO_2}} = \exp \left( \sum_k^{N_{\text{ionic}}} c_k \left( h_{s,k} + h_{g,\ce{CO_2}} \right) \right),
\end{equation}
where $k$ on the right-hand-side refers to the index of the ionic species and $N_{\text{ionic}}$ is the total number of ionic species.

\paragraph{Boundary conditions}
The boundary conditions for species fluxes at the \ce{Ag}/electrolyte interface ($y = 0$) are:
\begin{equation}
    \begin{split}
        \mathbf{J}_{ \ce{CO_2} } \cdot \mathbf{n} = \frac{i_{\ce{CO}, \text{local}} }{2 F}, \\
        \mathbf{J}_{ \ce{OH^-} } \cdot \mathbf{n} = - \frac{i_{ \ce{CO}, \text{local} } + i_{ \ce{H_2}, \text{local} }}{F}, \\
        \mathbf{J}_{ \text{all other species} } \cdot \mathbf{n} = 0.\\
    \end{split}
\end{equation}
$\mathbf{n}$ is the outward normal vector.

\section{\texorpdfstring{\ce{Ag}}{Ag} surface chemistry validation of implementation}\label{sec:Ag_surface_chemistry_validation}

\begin{figure}[!h]
    \centering
\includegraphics[width=0.5\textwidth]{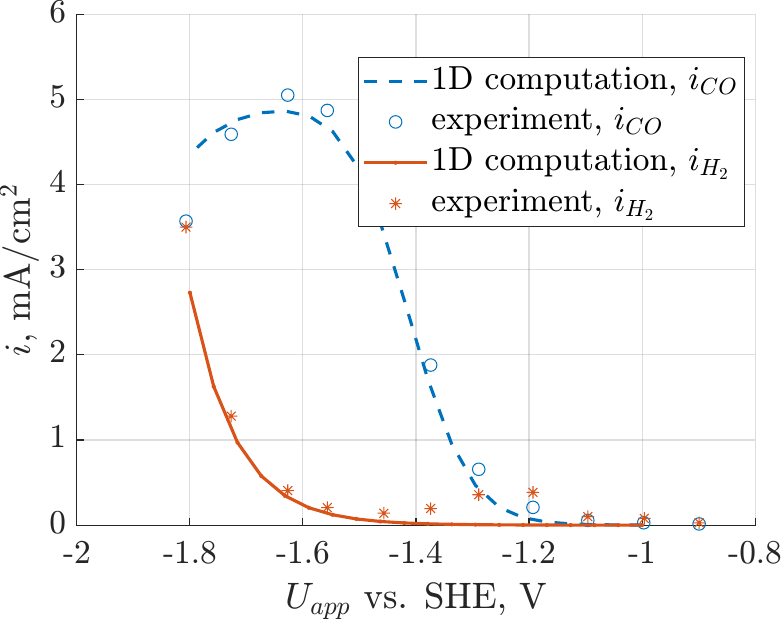}
    \caption{\doublespacing Comparison of 1D computations (from this current investigation) with experimental data \citep{clark2019influence}, for partial current densities (as given by the Tafel expressions in \cref{eq:i_CO_Tafel} and \cref{eq:i_H2_Tafel}) versus applied voltage $U_{app}$.}
\label{fig:1D_validation_bicarb_Ag}
\end{figure}

To ensure the \ce{Ag} surface chemistry is correctly implemented and compares well with the reference experimental observations, a one-dimensional (1D) computational case is run. For the computational setup, at one end the catalyst boundary \ce{Ag} is set; at the other end, the same bulk concentrations are used as is for all cases run in this study (as specified in section 2.2.1 of the main text).
The domain size is set by the boundary layer thickness, which is 174 $\mu$m as set by the experimental dataset used (from \citet{clark2019influence}, and specified as dataset 002 from \citet{corpus2023coupling}).

Results are shown in \cref{fig:1D_validation_bicarb_Ag}. The profiles for both $i_{\ce{CO}}$ and $i_{\ce{CO}}$ match well with the experimental results, especially the behavior at more negative $U_{app}$ values. The slight bump increase in $i_{H_2}$ at intermediate $U_{app}$ is not captured; this behavior has been attributed to the reduction of protons from bicarbonate \citep{koshy2021chemical}, which is an effect not considered in the Tafel modeling method utilized here. However, overall agreement in the trend and numerical values is acceptable and gives confidence that the important and relevant \ce{Ag} surface chemistry behavior is being captured.

\clearpage
\section{\texorpdfstring{\ce{Cu}}{Cu} surface chemistry formulation}\label{sec:Cu_surface_chemistry_formulation}
The Tafel expressions\citep{li2021electrokinetic} describe the \ce{Cu} surface reactions (given by eq. (14) through eq. (17) in the main text).
The Tafel expressions are re-written here:
\begin{equation}\label{eq:i_C2H4_Tafel}
\begin{split}
    i_{ \ce{C_2H_4}, \text{local} } = \left( \frac{c_{\ce{CO}}}{c_{ref}} \right) i_{0, \ce{C_2H_4} } 
    \left( - exp\left[ - \frac{\alpha_{c, \ce{C_2H_4} } F}{RT} \eta_{s, \ce{C_2H_4} } \right] \right),
\end{split}
\end{equation}
\begin{equation}\label{eq:i_C2H6O_Tafel}
\begin{split}
    i_{ \ce{C_2H_6O}, \text{local} } = \left( \frac{c_{ \ce{CO} }}{c_{ref}} \right) i_{0, \ce{C_2H_6O} } \left( - exp\left[ - \frac{\alpha_{c, \ce{C_2H_6O} } F}{RT} \eta_{s, \ce{C_2H_6O} } \right] \right),
\end{split}
\end{equation}
\begin{equation}\label{eq:i_CH4_Tafel}
\begin{split}
    i_{ \ce{CH_4}, \text{local} } = \left( \frac{c_{ \ce{CO} }}{c_{ref}} \right) i_{0, \ce{CH_4} } \left( - exp\left[ - \frac{\alpha_{c, \ce{CH_4} } F}{RT} \eta_{s, \ce{CH_4} } \right] \right),
\end{split}
\end{equation}
\begin{equation}\label{eq:i_H2_Tafel_Cu}
\begin{split}
    i_{ \ce{H_2}, \text{local} } = i_{0,\ce{H_2}} \left( - exp\left[ - \frac{\alpha_{c,\ce{H_2}} F}{RT} \eta_{s,\ce{H_2}} \right] \right).
\end{split}
\end{equation}
$c_{ref}$ for \cref{eq:i_C2H4_Tafel} through \cref{eq:i_H2_Tafel_Cu} is set to a value of $1$ mol/m$^3$.

The overpotential $U_{k}$ for each product species $k$ is written as
\begin{equation}
    \eta_{s, \ce{k}} = \phi_s - \phi_l - U_{ \ce{k} }^0, 
\end{equation}
for $\ce{k} = \{ \ce{C_2H_4}, \ce{C_2H_6O}, \ce{CH_4}, \ce{H_2} \}$. The necessary parameters $\alpha_{c,\ce{k}}$ are obtained by fitting to data provided by \citet{li2021electrokinetic} (specifically the pH $= 7.2$ set of data), yielding the following values of the transfer coefficient $\alpha_{c,\ce{k}}$ and exchange current density $i_{0,\ce{k}}$:
\begin{equation}
    \begin{split}
        \alpha_{c, \ce{C_2H_4} } = 0.4990, \\
        i_{0, \ce{C_2H_4} }  = 1.3835 \times 10^{-11} \text{A m}^{-2}, \\
        \alpha_{c, \ce{C_2H_6O} } = 0.4452, \\
        i_{0, \ce{C_2H_6O} }  = 8.9302 \times 10^{-11} \text{A m}^{-2}, \\
        \alpha_{c, \ce{CH_4} } = 0.4731, \\
        i_{0, \ce{CH_4} }  = 2.0064 \times 10^{-11} \text{A m}^{-2}, \\
        \alpha_{c, \ce{H_2} } = 0.2547, \\
        i_{0, \ce{H_2} }  = 4.627 \times 10^{-5} \text{A m}^{-2}. \\
    \end{split}
\end{equation}
For these parameters reported here, the reference voltage $U_{\ce{k}}^0 = 0.0$V vs. SHE for \ce{k} = \ce{C_2H_4}, \ce{C_2H_6O}, \ce{CH_4}, and \ce{H_2}.

\paragraph{Boundary conditions}
The boundary conditions for species fluxes at the \ce{Cu}/electrolyte interface ($y = 0$) are:
\begin{equation}
    \begin{split}
        \mathbf{J}_{ \ce{CO} } \cdot \mathbf{n} =  \frac{i_{ \ce{C_2H_4}, \text{local} } }{4F} + \frac{i_{ \ce{C_2H_6O}, \text{local} } }{4F} + \frac{i_{ \ce{CH_4}, \text{local} } }{6F}, \\
        \mathbf{J}_{ \ce{OH^-} } \cdot \mathbf{n} = - \frac{ i_{ \ce{C_2H_4}, \text{local} } + i_{ \ce{C_2H_6O}, \text{local} } + i_{ \ce{CH_4}, \text{local} } + i_{\ce{H_2}, \text{local}} }{F}, \\
        \mathbf{J}_{ \text{all other species} } \cdot \mathbf{n} = 0.\\
    \end{split}
\end{equation}
$\mathbf{n}$ is the outward normal vector.

\section{\texorpdfstring{\ce{Cu}}{Cu} surface chemistry validation}\label{sec:Cu_surface_chemistry_validation}

\begin{figure}[!h]
    \centering
    \begin{subfigure}[b]{0.48\textwidth}
    \centering
        \includegraphics[width=\textwidth]{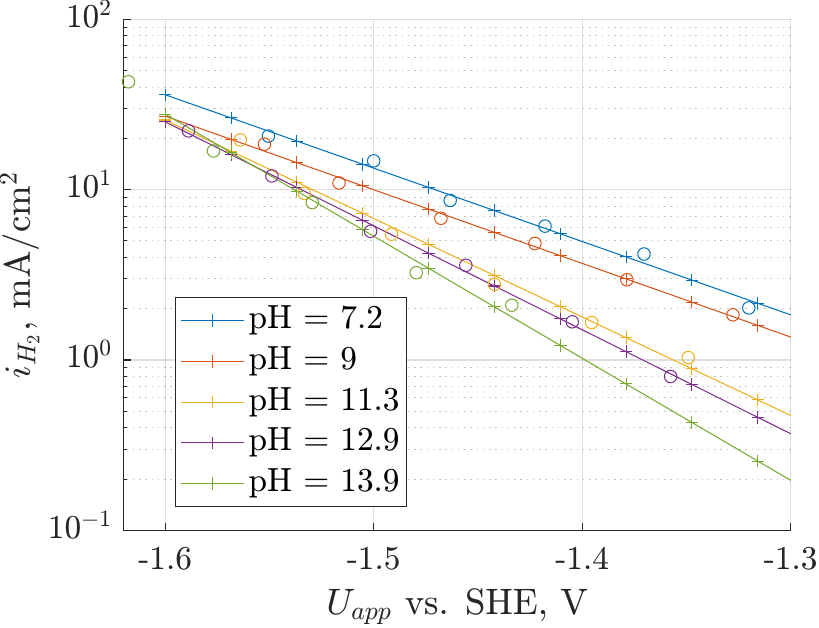}
        \caption{\doublespacing \ce{H_2} (as given by the Tafel expression in \cref{eq:i_H2_Tafel_Cu})}
    \label{fig:1D_validation_Cu_H2}
    \end{subfigure}
    \begin{subfigure}[b]{0.48\textwidth}
        \centering
        \includegraphics[width=\textwidth]{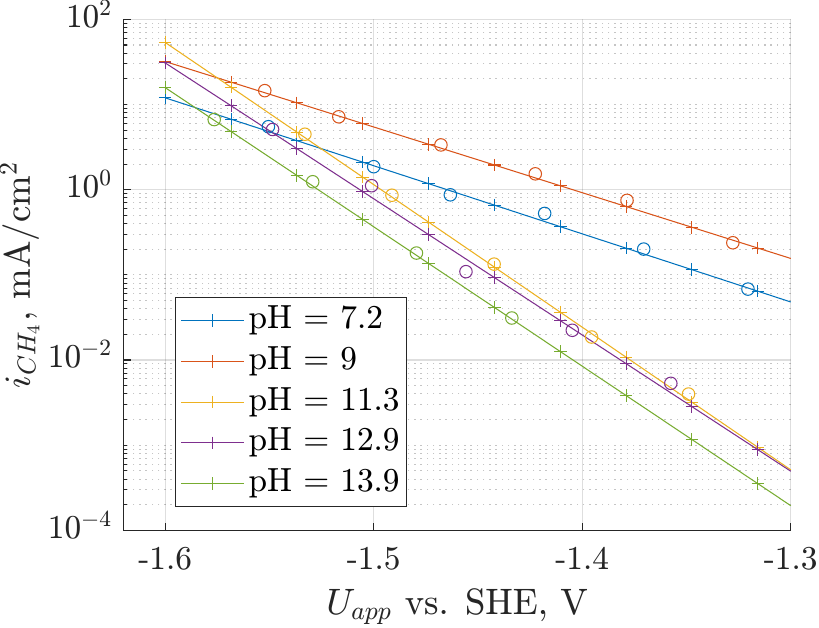}
        \caption{\doublespacing \ce{CH_4} (as given by the Tafel expression in \cref{eq:i_CH4_Tafel})}
    \label{fig:1D_validation_Cu_CH4}
    \end{subfigure}
    \caption{\doublespacing Comparison of 1D computations (from this current investigation) with experimental data \citep{li2021electrokinetic}, for partial current densities (as given by the Tafel expression in \cref{eq:i_H2_Tafel_Cu}) versus applied voltage $U_{app}$. Experimental values are shown with empty circles.}
    \label{fig:1D_validation_Cu_H2_CH4}
\end{figure}

\begin{figure}[!h]
    \centering
\includegraphics[width=0.5\textwidth]{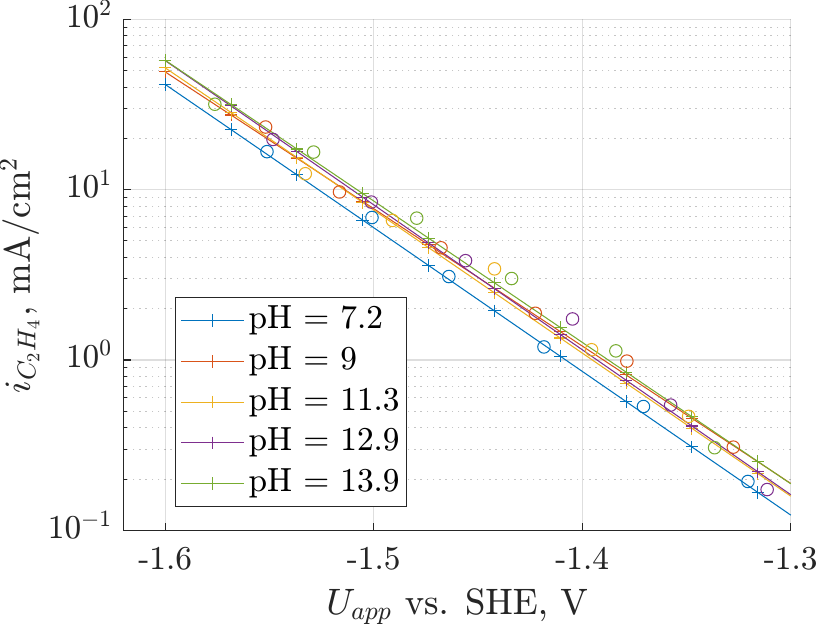}
    \caption{\doublespacing Comparison of 1D computations (from this current investigation) with experimental data \citep{li2021electrokinetic}, for partial current densities (as given by the Tafel expression in \cref{eq:i_C2H4_Tafel}) versus applied voltage $U_{app}$. Experimental values are shown with empty circles.}
\label{fig:1D_validation_Cu_C2H4}
\end{figure}

To ensure the \ce{Cu} surface chemistry is correctly implemented and compares well with the reference experimental observations, a one-dimensional (1D) computational case is run. The 1D computational setup follows that of the 1D \ce{Ag} case in \cref{sec:Ag_surface_chemistry_validation}, with the sole difference being that the bulk \ce{CO} concentration is set to $0.95$ mM (corresponding to the solubility limit of \ce{CO} in water at a pressure of $1$ atmosphere). Each of the $5$ reported pH values in \citet{li2021electrokinetic} yields a different set of Tafel parameter values. Validation results for all $5$ pH values are shown here in order to check the implementation; however, as discussed in \cref{sec:Cu_surface_chemistry_formulation}, only the parameters for pH $= 7.2$ are used in computing the results of the current study. A very thin boundary layer thickness of $0.01$ $\mu$m is set to ensure the simulations exist comfortably in the mass transport-limited regime. The results for three product species -- \ce{H_2}, \ce{CH_4}, \ce{C2H_4} -- are shown in \cref{fig:1D_validation_Cu_H2}, \cref{fig:1D_validation_Cu_CH4}, and \cref{fig:1D_validation_Cu_C2H4}. This validation exercise checks that the implemented Tafel expressions yield current density profiles that match well with the corresponding experimental values; this is indeed the case.

\clearpage
\section{Manually-patterned \texorpdfstring{$N = 2$}{N = 2} cases, Faradaic Efficiency (FE)}\label{sec:Manual_optimization_FE}

\begin{figure}[!h]
    \centering
    \includegraphics[width=\textwidth]{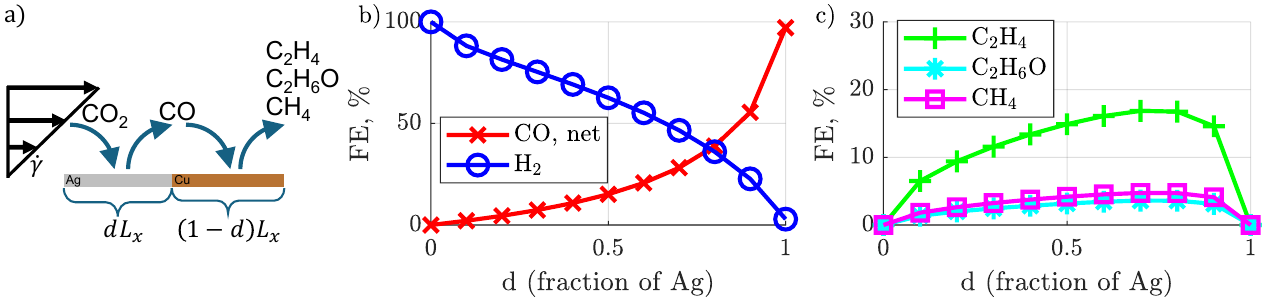}
    \caption{\doublespacing a) A schematic depicting the patterning configuration $N = 2$, along with the cascade reaction pathway considered: \ce{ CO_2 -> CO -> \{C_2H_4, C_2H_6O, CH_4\} }. $d$ denotes the fraction of the electrode length assigned to the $\ce{Ag}$ section. b) Plots of Faradaic efficiency (FE), as a percentage, for net \ce{CO} and for \ce{H_2} as a function of $d$. c) Plots of FE, as a percentage, for \ce{C_2H_4}, \ce{C_2H_6O}, and \ce{CH_4} as a function of $d$. The cases run to obtain the current density values in b and c use the same conditions as chosen in fig. 1 of the main text: flow rate = $3.0$ ml/min, $U_{app} = -1.7$ V vs. SHE.}
    \label{fig:trivial_opt_figure_FE}
\end{figure}

The Faradaic efficiency (FE) for product species $k$ is computed as
\begin{equation}
    \text{FE}_k = \frac{i_k}{i_{\ce{H_2}} + i_{\ce{CO}, \mathrm{net}} + i_{\ce{C_2H_4}} + i_{\ce{C_2H_6O}} + i_{\ce{CH_4}}}.
\end{equation}

 In \cref{fig:trivial_opt_figure_FE}, it is observed that a trivial optimization maximum occurs for FE of \ce{CO} (at $d = 1$) and for \ce{H_2} (at $d = 0$), and a nontrivial optimization maximum occurs for FE for \ce{C_2H_4}, \ce{C_2H_6O}, and \ce{CH_4}, just as was observed for the current density values as shown in section 3.1 of the main text. 
 However, a key difference is that the maximum FE values for the nontrivial optimization products occur at a value of $d \simeq 0.8$, which is shifted significantly towards favoring \ce{Ag} compared to the maximum at $d \simeq 0.5$ for the $i$ values. Because the \ce{Cu} kinetics used in the current study has a strong propensity towards producing \ce{H_2}, the maximum FE for the other \ce{Cu} products is achieved by using configurations that use a smaller proportion of \ce{Cu} to minimize the relative magnitude of the \ce{H_2} production.

\section{Average surface concentrations of \texorpdfstring{\ce{CO_2}, \ce{CO}, \ce{C_2H_4}}{CO_2, CO, C_2H_4}}\label{sec:average_surface_concs_CO2_CO_C2H4}
\begin{figure}[H]
    \centering
    \begin{subfigure}[t]{\textwidth}
        \centering
        \includegraphics[width=\textwidth]{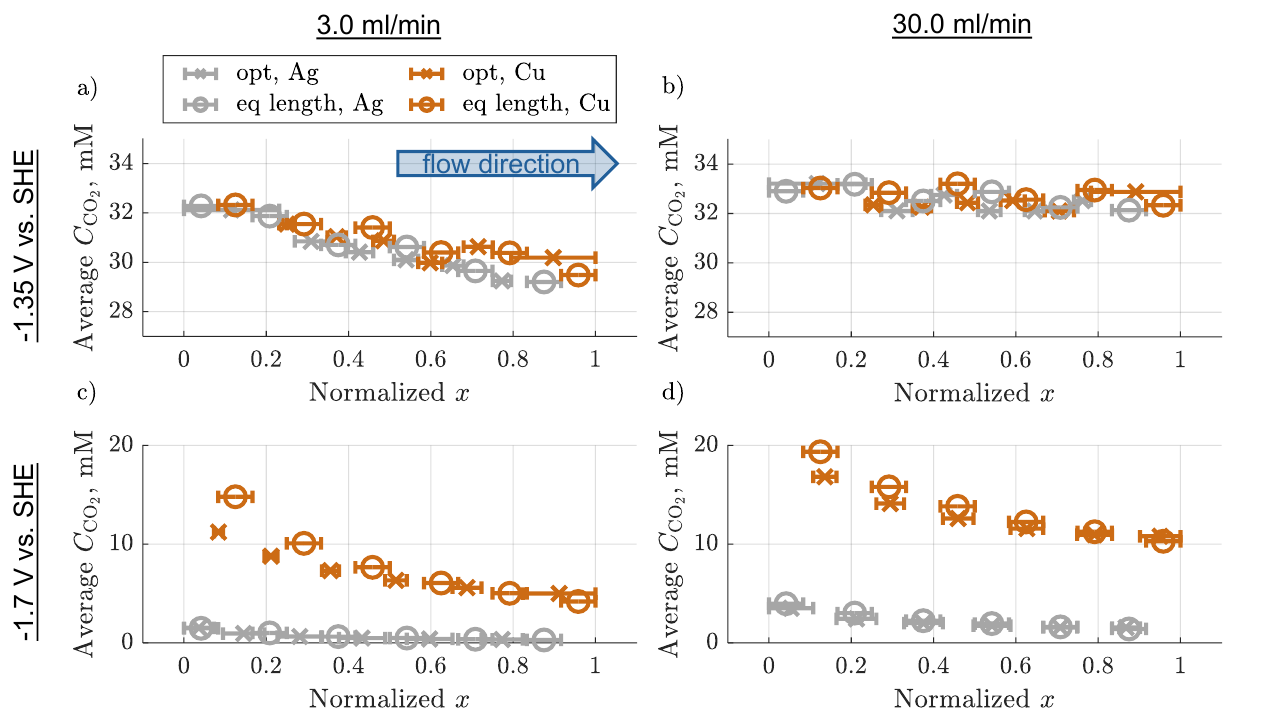}
        \caption{\doublespacing \ce{CO_2} plots.}
        \label{fig:avg_concs_CO2_plot}
    \end{subfigure}
    \vspace{-0.5em} 

    \begin{subfigure}[t]{\textwidth}
        \centering
        \includegraphics[width=\textwidth]{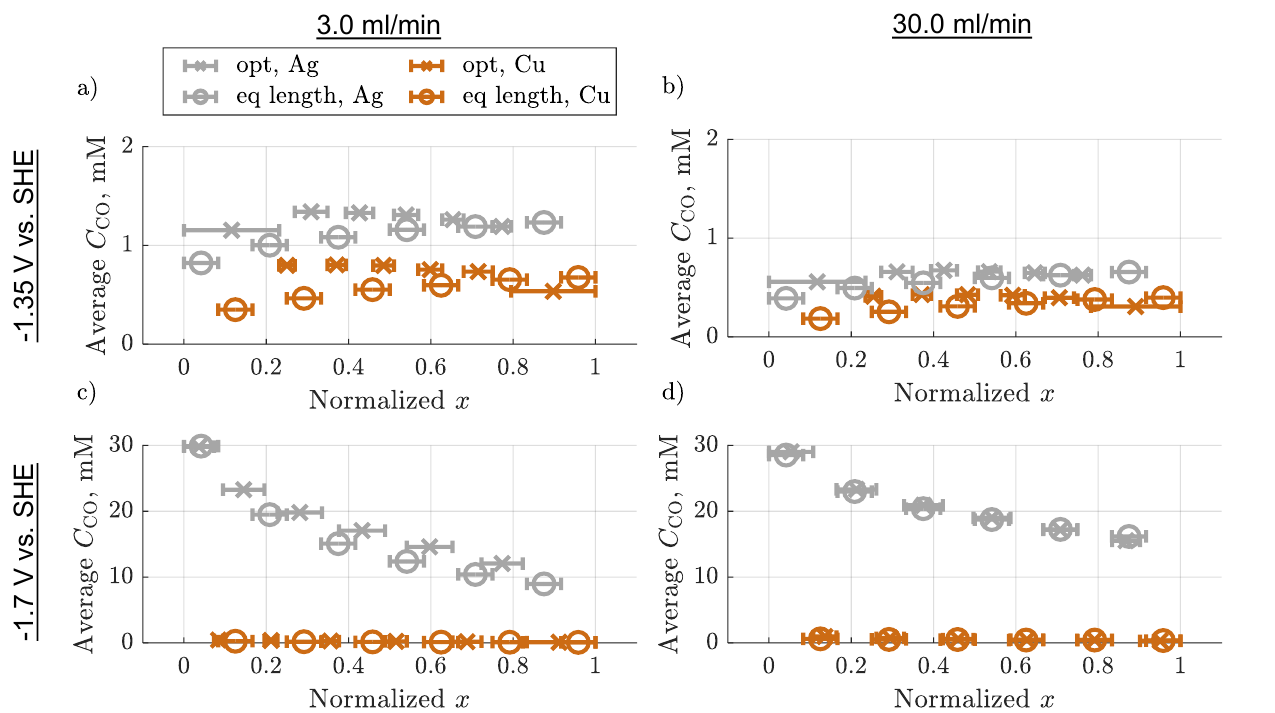}
        \caption{\doublespacing \ce{CO} plots.}
        \label{fig:avg_concs_CO_plot}
    \end{subfigure}
    \caption{\doublespacing \ce{CO_2} and \ce{CO} average surface concentrations, for $N = 12$, on each section.}
    \label{fig:avg_concs_combined_CO_2_CO}
\end{figure}
\begin{figure}[!h]
    \centering
    \includegraphics[width=\textwidth]{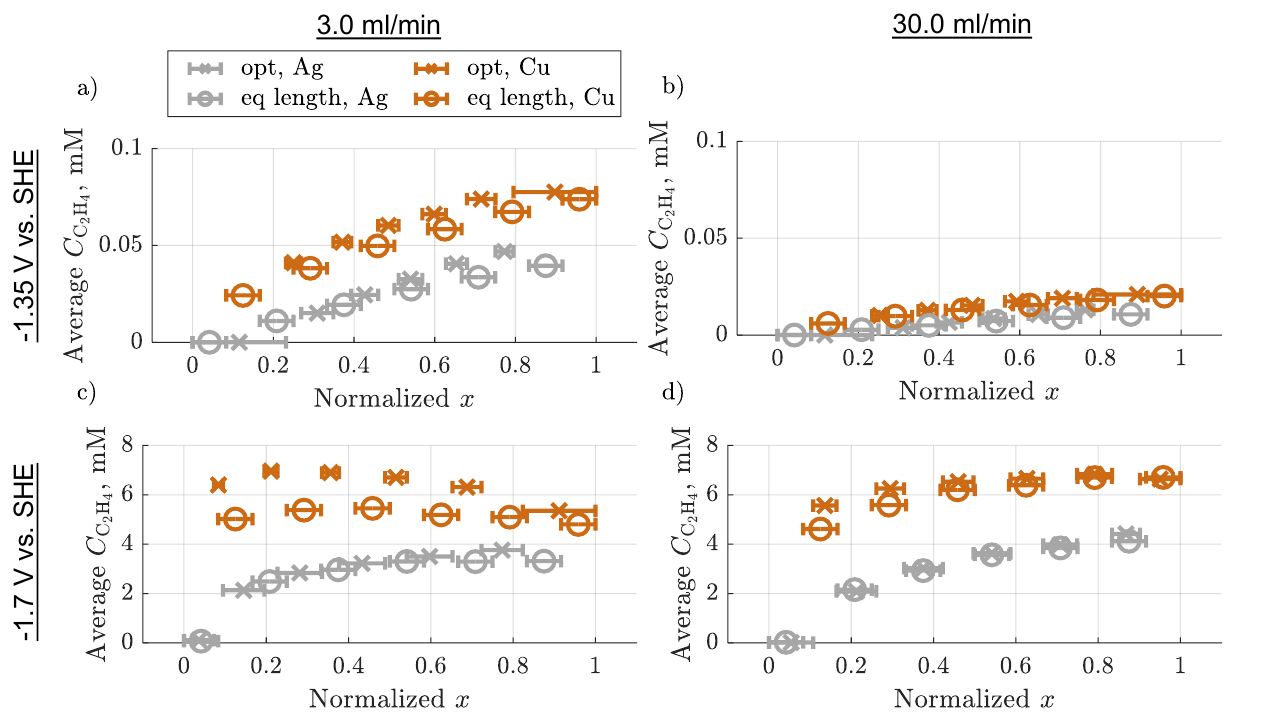}
    \caption{\doublespacing \ce{C_2H_4} average surface concentrations, for $N = 12$, on each section.}
    \label{fig:avg_concs_C2H4_plot}
\end{figure}

As discussed in the analysis for section 3.4.2 of the main text, \cref{fig:avg_concs_combined_CO_2_CO} and \cref{fig:avg_concs_C2H4_plot} provide the average surface concentration of \ce{CO_2}, \ce{CO}, and \ce{C_2H_4} on each of the \ce{Ag} and \ce{Cu} sections for the $N = 12$ equal length and optimized cases.

\clearpage

\begingroup
\bibliographystyle{unsrtnat}
\bibliography{references}
\endgroup